%% file: motor_methods_main.tex
\documentclass[12pt]{article}
\usepackage{graphicx}
\usepackage{amsmath}
\usepackage{amssymb}
\usepackage{mathtools}
\usepackage{bm}
\usepackage{xr}
\usepackage{xcite}
\usepackage{cite}
\usepackage{mwe}
\usepackage{times}
\usepackage{caption}
\usepackage{ragged2e}
\usepackage{array}
\usepackage{color}
\usepackage{float}
\usepackage{xparse}
\usepackage{chemformula}
\usepackage{silence}
\usepackage{setspace}
\usepackage[toc,page]{appendix}

  \newcounter{lastnote}

\newcommand{\bff}{{\bf f}}

\newcommand{\bF}{{\bf F}}
\newcommand{\bT}{{\bf T}}

\newcommand{\br}{{\bf r}}

\newcommand{\bh}{{\bf h}}

\newcommand{\rsqr}{r_{i,j}^2}
\newcommand{\del}{\partial}
\newcommand{\kb}{k_{\textrm{B}}}
\newcommand{\para}{\parallel}
\newcommand{\Fs}{f_{{\rm stall}}}
\newcommand{\hs}{h_{{\rm stall}}}

\newcommand{\Ds}{\Delta{s}}
\newcommand{\Dt}{\Delta t}
\newcommand{\uhat}{{\hat{u}}}
\newcommand{\pder}[2][]{\frac{\partial#1}{\partial#2}}
\newcommand{\pdertwo}[2][]{\frac{\partial^2#1}{\partial#2^2}}

\newcommand{\psij}{\psi_{i,j}}
\newcommand{\hij}{h_{i,j}}

\newcommand{\Ke}{K_{\rm E}}

\newcommand{\kcl}{k_{\rm cl}}
\newcommand{\hcl}{h_{\rm cl}}
\newcommand{\Ucl}{U_{i,j}}

\newcommand{\micron}{$\mu$m}

\newcommand{\Vbd}{V_{\rm bind}}

\DeclareMathOperator{\sgn}{sgn}
\DeclareMathOperator{\CDF}{{\rm CDF}}
\DeclareMathOperator{\PDF}{{\rm PDF}}
\DeclareMathOperator\erf{erf}
\NewDocumentCommand{\kon}{O{}}{k_{{\rm on} #1}}
\NewDocumentCommand{\koff}{O{}}{k_{{\rm off} #1}}
\NewDocumentCommand{\ko}{O{}}{k_{{\rm o} #1}}
\NewDocumentCommand{\PSI}{O{m} O{n} O{k}}{\psi_{i,j}^{#1, #2, #3}}
\NewDocumentCommand{\Dr}{O{i,j}}{{\bf r}_{#1}}
\NewDocumentCommand{\cb}{O{i,j}}{\uhat_i \cdot \uhat_j}
\NewDocumentCommand{\ru}{O{i} O{j}}{{\bf r}_{#1, #2}\cdot\uhat_{#1}}
\NewDocumentCommand{\q}{O{i,j}}{q_{#1}}

\title{Comparison of explicit and mean-field models of  cytoskeletal filaments with crosslinking motors}

\author{Adam R. Lamson$^{1}$, Jeffrey M. Moore$^{1}$, Fang Fang$^2$, Matthew A. Glaser$^{1}$,\\
  Michael Shelley$^{2,3}$, Meredith D. Betterton$^{1}$\\
  \normalsize{$^{1}$Department of Physics, University of Colorado Boulder}\\
   \normalsize{$^{2}$Courant Institute, New York University}\\
   \normalsize{$^{3}$Center for Computational Biology, Flatiron Institute}\\
}

\date{}

\begin{document}
\maketitle

\newpage

\begin{abstract}
   In cells, cytoskeletal filament networks are responsible for cell movement,
   growth, and division.  Filaments in the cytoskeleton are driven and
   organized by crosslinking molecular motors. In reconstituted cytoskeletal
   systems, motor activity is responsible for far-from-equilibrium phenomena
   such as active stress, self-organized flow, and spontaneous nematic defect
   generation.  How microscopic interactions between motors and filaments lead
   to larger-scale dynamics remains incompletely understood.  To build from
   motor-filament interactions to predict bulk behavior of cytoskeletal
   systems, more computationally efficient techniques for modeling
   motor-filament interactions are needed. Here we derive a coarse-graining
   hierarchy of explicit and continuum models for crosslinking motors that bind
   to and walk on filament pairs. We compare the steady-state motor
   distribution and motor-induced filament motion for the different models and
   analyze their computational cost.  All three models agree well in the limit
   of fast motor binding kinetics. Evolving a truncated moment expansion of
   motor density speeds the computation by $10^3$--$10^6$ compared to the
   explicit or continuous-density simulations, suggesting an approach for more
   efficient simulation of large networks.  These  tools facilitate further
   study of motor-filament networks on micrometer to millimeter length scales. 
 \end{abstract}

\section{Introduction}%
\label{sec:introduction}
The cytoskeleton generates force and reorganizes to perform important cellular
processes~\cite{howard01}, including cell motility~\cite{bray01, blanchoin14},
cytokinesis~\cite{pollard09}, and chromosome segregation in
mitosis~\cite{mcintosh12}.  The cytoskeleton is made of polymer filaments,
molecular motors, and associated proteins. The two best-studied cytoskeletal
filaments are actin and microtubules~\cite{howard01}. It remains incompletely
understood how diverse cytoskeletal structures dynamically assemble and
generate force 
of pN to nN~\cite{howard01, bray01}.

Force generation and reorganization in the cytoskeleton depend on the activity
of crosslinking motor proteins that align and slide pairs of filaments (Figure
1).  Reorganization of actin networks by myosin motors is important for muscle
contraction~\cite{huxley57,huxley96, huxley97}, cell crawling and shape
change~\cite{gupton06, fournier10, barnhart11}, and
cytokinesis~\cite{laevsky03, pollard09}.  Microtubule  sliding by crosslinking
kinesin and dynein motors contributes to mitotic spindle
assembly~\cite{hagan92, saunders95, kapoor00, cai08, mcintosh12}, chromosome
segregation~\cite{wollman08, civelekoglu-scholey10a, she17, vukusic17},
cytoplasmic stirring in Drosophila oocytes~\cite{ganguly12}, and beating of
cilia and flagella~\cite{satir68, summers71, king18}.  

Filament-motor interactions  produce diverse cellular structures and dynamics,
but linking molecular properties of motors to larger-scale assembly behavior
remains challenging. Crosslinking motors  vary in binding affinity, speed,
processivity, and force-velocity relation. These same ingredients can be
reconstituted and show dynamic self-organization into asters or contractile
bundles~\cite{nedelec97, surrey01, Backouche2006}, active liquid
crystals~\cite{sanchez12a, doostmohammadi18, lemma19, duclos20}, or other
structures~\cite{brugues12, roostalu18, weirich19}. Even in reconstituted
systems, our ability to predict and control dynamics and self-organization is
limited.

Improved theory and simulation of cytoskeletal assemblies with crosslinking
motors would allow better prediction of both cellular and reconstituted
systems. Currently few mesoscale modeling methods for filament-motor systems
are available between explicit particle simulations and continuum hydrodynamic
theory.  Explicit motor simulations have several existing software tools,
including Cytosim~\cite{nedelec07}, MEDYAN~\cite{popov16}, and
AFINES~\cite{freedman17}, and others~\cite{head14}.  Explicit motor simulations
are straightforward to extend to include, for example,  a new force-velocity
relation or motor cooperativity. However, the cost of explicit particle
simulations scales linearly or quadratically with the number of particles
(depending on the type of interactions), making simulation of large systems
challenging.  Continuum models of coarse-grained fields can be computationally
tractable and  predict macroscopic behavior~\cite{aranson05, kruse05,
  saintillan08, giomi13, gao15, white15b, maryshev18, furthauer19}.  Current
continuum models invoke symmetry considerations to determine the structure of
the model without reference to an underlying microscopic
mechanisms~\cite{aranson05, ziebert07, Giomi2013, Lenz2014}, or simplify a
microscopic model by making assumptions about the physics of
motor~\cite{Kruse2000, Kruse2005, ahmadi05, ahmadi06, Liverpool2005,
  swaminathan11, gao15, gao15a, gao17}.  Furthermore, previous continuum
theories have  coarse-grained the filament distribution, with simplifying
assumptions about the motor distribution. This presents an opportunity to
better understand how the distribution of motors evolves and affects filament
motion. Further development of mesoscale modeling techniques focusing on
crosslinking motors could help bridge the gap between detailed explicit
particle models and continuum theories.

To develop mesoscale modeling tools, we focus on the fundamental unit of a
crosslinked filament network: two filaments with crosslinking motors
that  translate and rotate the
filaments. We study three different model representations in a coarse-graining
hierarchy and compare computational cost and accuracy. For explicit motors, we
extend previous work that uses Brownian dynamics and kinetic Monte Carlo
simulation to handle filament motion and binding kinetics~\cite{gao15, gao15a,
  blackwell16, blackwell17a, rincon17, lamson19, edelmaier20}.  At the first
level of coarse-graining, we average over discrete bound motors to compute the
continuum mean-field motor density (MFMD) between filaments, and evolve this
density according to a first-order Fokker-Planck equation~\cite{blackwell16}.
This requires computing the solution to a single partial differential equation
(PDE) for each filament pair,  rather than separately tracking each individual
motor. The MFMD determines the force and torque  on each filament needed to
evolve its position and orientation.   At the second level of coarse-graining,
we expand the MFMD in moments to derive a system of ordinary differential
equations (ODEs) for the time evolution of the moments.  While the moment
expansion does not close, an approximate treatment of filament motion can be
modeled by low-order moments. To compare these three model implementations, we
consider test cases of parallel, antiparallel, and perpendicular filaments.
Under the same initial conditions, the three model implementations give similar
results on average.  Remarkably, the reduced moment expansion achieves a
computational cost that is $10^3$--$10^6$ lower than the other models,
suggesting a route to computationally tractable large-scale simulations.

\section{Model overview}%
\label{sec:filament_network_overview}
\begin{figure}[htpb]
  \centering
  \includegraphics[width=1.0\linewidth]{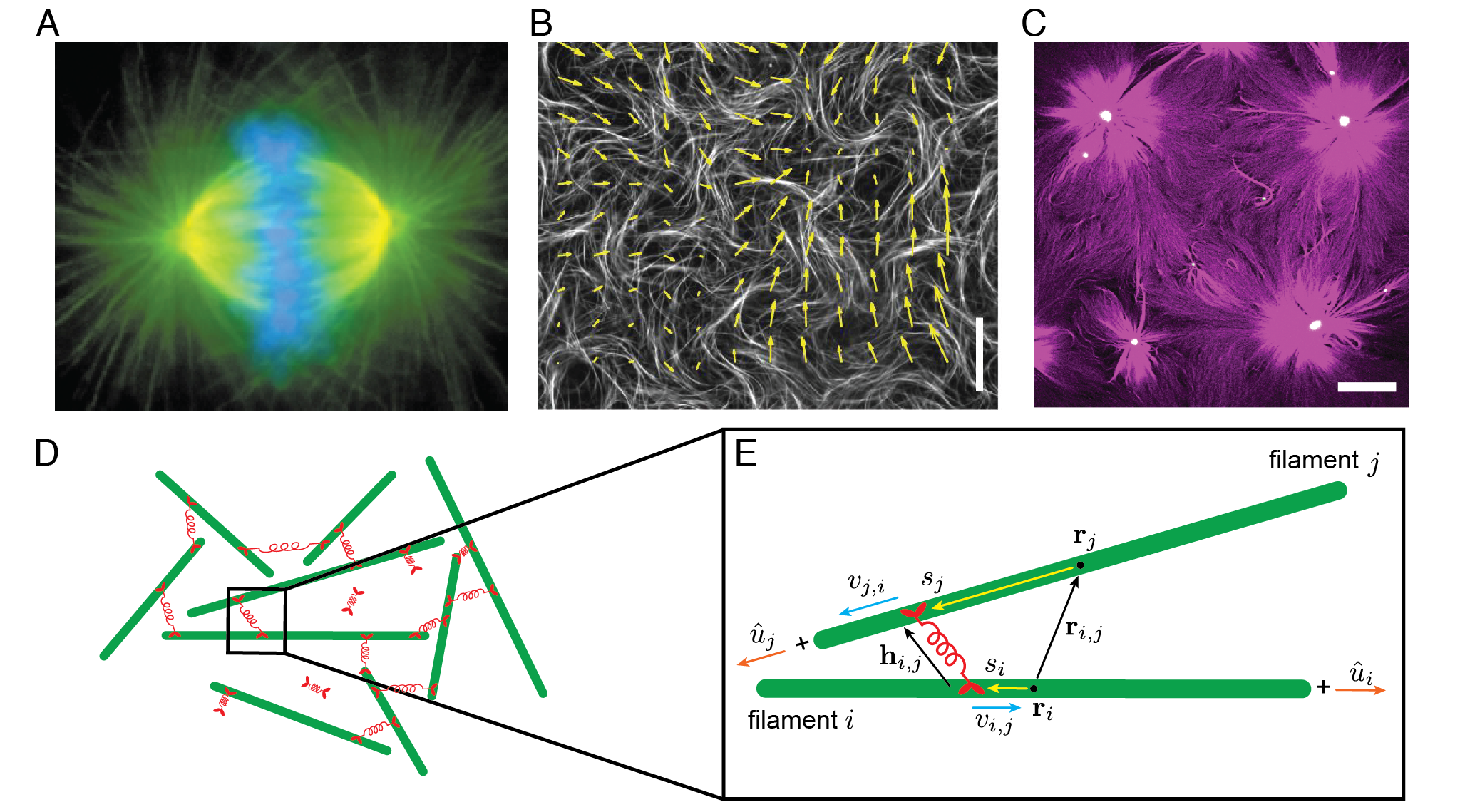}
  \caption{\footnotesize{Experimental systems of cytoskeletal filaments  with
      crosslinking motors and overview of the model. {\bf A-C} Fluoresence
      microscopy images of cytoskeletal networks.  {\bf A} Mitotic spindle
      showing microtubules~(green), chromosomes~(blue), and spindle-pole
      component TPX2~(red)~\cite{wittmann01}.  {\bf B} Reconstituted active gel
      of microtubules~(white) driven by crosslinking kinesin motor clusters
      with local flow field shown~(yellow arrows)~\cite{sanchez12a}. Scale bar:
      80 \micron. {\bf C} Reconstituted active network of actin~(magenta) and
      myosin-II~(green)~\cite{wollrab19}.  Scale bar: 50 \micron.  {\bf D}
      Schematic of filament-motor network with green filaments and red motors.
      {\bf E} Schematic of filament pair~(green) crosslinked by a motor~(red)
      with model variables position of filament $i$'s center $\br_i$,
      orientation vector of filament $i$ $\uhat_i$, vector between filament
      centers $\br_{i,j} = \br_j - \br_i$, vector between motor heads
      $\bh_{i,j}$, motor tether extension $|\hij|$, and motor speed on filament
      $i$ while attached to filament $j$}.}%
  \label{fig:1_overview}
\end{figure}

We consider a pair of rigid, inextensible filaments that move and reorient
under the force and torque applied by crosslinking motors. Filaments move in
three dimensions, experience viscous drag, and are constrained to prevent
overlaps.  Motors bind to and unbind from the filaments consistent with
detailed balance in binding. Crosslinking motors walk with a force-dependent
velocity toward filament plus ends and unbind when they reach the ends. We
investigate models at three  levels: an explicit
motor model where motors are represented with a discrete density,  a continuum
mean-field motor density (MFMD) model, and  a moment expansion model. 

\subsection{Filaments}%
\label{sub:filament_model}
We model filament motion using Brownian dynamics, balancing the force applied
by motors against viscous drag and constraint forces. Because the force that
induces Brownian motion is typically smaller than that due to motors, we
neglect Brownian noise~\cite{brangwynne08}.

Filaments translate according to
the force-balance equation 
\begin{equation}
  \label{eq:dot_r_i}
  \dot{\br}_{i} = {\bf{M}}_i \left(\sum_n \bF_{n,i} \right),
\end{equation}
where $\br_i$ is the center of filament $i$ with mobility matrix ${\bf M}_i$
acted on by forces $\bF_{n,i}$.  The mobility matrix for a perfectly rigid rod
in a viscous medium is
\begin{equation}
  \label{eq:mob_mat}
  {\bf M}_i = \left(\left( \gamma_{\para, i}
    - \gamma_{\perp, i}\right)\uhat_i\uhat_i
  + \gamma_{\perp, i }{\bf I}\right)^{-1},
\end{equation}
where $\bf I$ is the identity matrix and $\gamma_{\para,i}$ and
$\gamma_{\perp,i}$ are the parallel and perpendicular drag coefficients with
respect to the filament orientation $\uhat_i$. Cytoskeletal filaments with
length $L_i$ and diameter $D_{{\rm fil}}$  typically have a large aspect ratio
$L_i/D_{{\rm fil}} \gg 1$, so we approximate the drag coefficients using
slender body theory~\cite{tao05}.

The torque-balance equation is 
\begin{equation}
  \label{eq:dot_uhat_i}
  \dot{\uhat}_i = \frac{1}{\gamma_{\theta,i}}
  \left(\sum_n \bT_{n,i} \right) \times \uhat_i,
\end{equation}
where $\bT_{n,i}$ are the torques acting on filament $i$ and $\gamma_{\theta,i}$
is the rotational drag coefficient about the center of filament $i$.

The force and torque exerted by crosslinking motors depend on
where motors are attached, the motor tether extension, and the relative
position and orientation of filaments. 
Given the crosslinking motor distribution along the filaments $\psij(s_i,s_j)$,
where $s_i$ is the bound motor head position on filament $i$, the total
crosslinking force and torque exerted by filament $i$ on filament $j$ are
\begin{eqnarray}
  \label{eq:F_mean}
  \bF_{i,j} &=& \int_{L_{i}} \int_{L_{j}} \bff_{i,j}(s_i, s_j )\psij(s_i, s_j)  ds_i ds_j,\\
  \label{eq:T_mean}
  \bT_{i,j} &=& \int_{L_{i}} \int_{L_{j}} s_j \uhat_j \times \bff_{i,j}(s_i, s_j)\psij(s_i, s_j) ds_i ds_j.
\end{eqnarray}
where $\bff_{i,j}(s_i, s_j)$ is the force exerted on filament $j$ by the
crosslinking motor attached at $s_i$ and $s_j$ (Figure~\ref{fig:1_overview}E).
For brevity, we use subscripts on variables such as $\bff_{i,j}$ to indicate
that these are functions of the relative position and orientation of filaments
$i$ and $j$.  Our three model implementations all models use equations
(\ref{eq:F_mean}) and (\ref{eq:T_mean}) to compute the force and torque that
evolve filament position and orientation but differ in how the computation of
$\psij$.

We constrain the motion of filaments to prevent overlap, which avoids numerical
instabilities introduced by a hard potential between filaments.  To implement
the constraint, we construct a vector $\uhat_{{\rm min}}$ that is perpendicular
to both infinite carrier lines defined by $\uhat_i$ and $\uhat_j$ and parallel
to the vector of closest approach between these lines.  The vector $\uhat_{{\rm
min}}$ is used to define two normal planes that confine the filaments,
leading to the modified force and torque
\begin{eqnarray}
  \label{eq:force_constraint}
  \widetilde{\bF}_{i,j} & = & \bF_{i,j} - (\bF_{i,j}\cdot\uhat_{{\rm min}})\uhat_{{\rm min}} \\
  \label{eq:torque_constraint}
  \widetilde{\bT}_{i,j} & = & (\bT_{i,j}\cdot\uhat_{{\rm min}})\uhat_{{\rm min}}.
\end{eqnarray}
Note that for filaments lying in the same confining plane and $|\hat{u}_{i}
\cdot \hat{u}_j| < 1$, $\uhat_{{\rm min}} = {\bf 0}$ and our constraints break
down. However, if only the first condition is satisfied, i.e., (anti)parallel
filaments, $\widetilde{\bT}_{i,j} = 0$ and $\widetilde{\bF}_{i,j}$ is parallel
to $\uhat_{i}$ and $\uhat_{j}$. After computing the force and torque, we
numerically integrate  equations (\ref{eq:dot_r_i}) and (\ref{eq:dot_uhat_i})
to update filament position and orientation.

\subsection{Motors}%
\label{sub:motors}
In our model motors bind and unbind, crosslink between two filaments, exert
force and torque when crosslinking, and walk with a force-dependent velocity.  Typically
motor proteins diffuse in solution until they are near a filament, then
stochastically bind to that filament. Once one head binds, the other head can
bind to a second filament, forming a crosslink, or the motor can unbind.
Crosslinking motors can unbind to a state with one head bound, or can unbind
completely from both filaments. 
We consider an infinite reservoir of unbound motor proteins.  The diffusion of
motors in solution is fast relative to the motion of filaments, so we assume
the motor reservoir has uniform, constant concentration.  We neglect steric
interactions between motors.  This approximation holds for filaments sparsely
populated with motors and motors that do not cluster on filaments or in
solution. 

Motors crosslinking filaments have a potential energy $\Ucl(s_i, s_j)$ (Figure
1). The energy depends on the motor head separation vector $\bh_{i,j}(s_i,s_j)
= \br_j + s_j\uhat_j - (\br_i + s_i\uhat_i)$ that gives the motor tether
extension
\begin{equation} 
  \label{eq:h_mag} 
  h_{i,j}(s_i, s_j) = \sqrt{r_{i,j}^2 + s_j^2 +
    s_i^2 + 2 \br_{i,j} \cdot (s_j\uhat_j - s_i\uhat_i) - 2 s_i s_j (\uhat_i
    \cdot \uhat_j)}, 
\end{equation} 
where $\br_{i,j} = \br_j - \br_i$ and $\rsqr = \Dr \cdot \Dr$
(Figure~\ref{fig:1_overview}E). 

The bound motor heads walk with a speed $v_{i,j}$ that depends on the force
component on the motor head parallel to the walking direction, $\uhat_i \cdot
\bff_{j,i}$~\cite{visscher99}. This projected force is used to determine the
motor speed via the force-velocity relation, as discussed below. 
This model is based on processive microtubule motors such as kinesin and
dynein, but a similar model has been used for myosin
minifilaments~\cite{popov16, freedman17}.

\begin{figure}[htpb]
  \centering
  \includegraphics[width=1.0\linewidth]{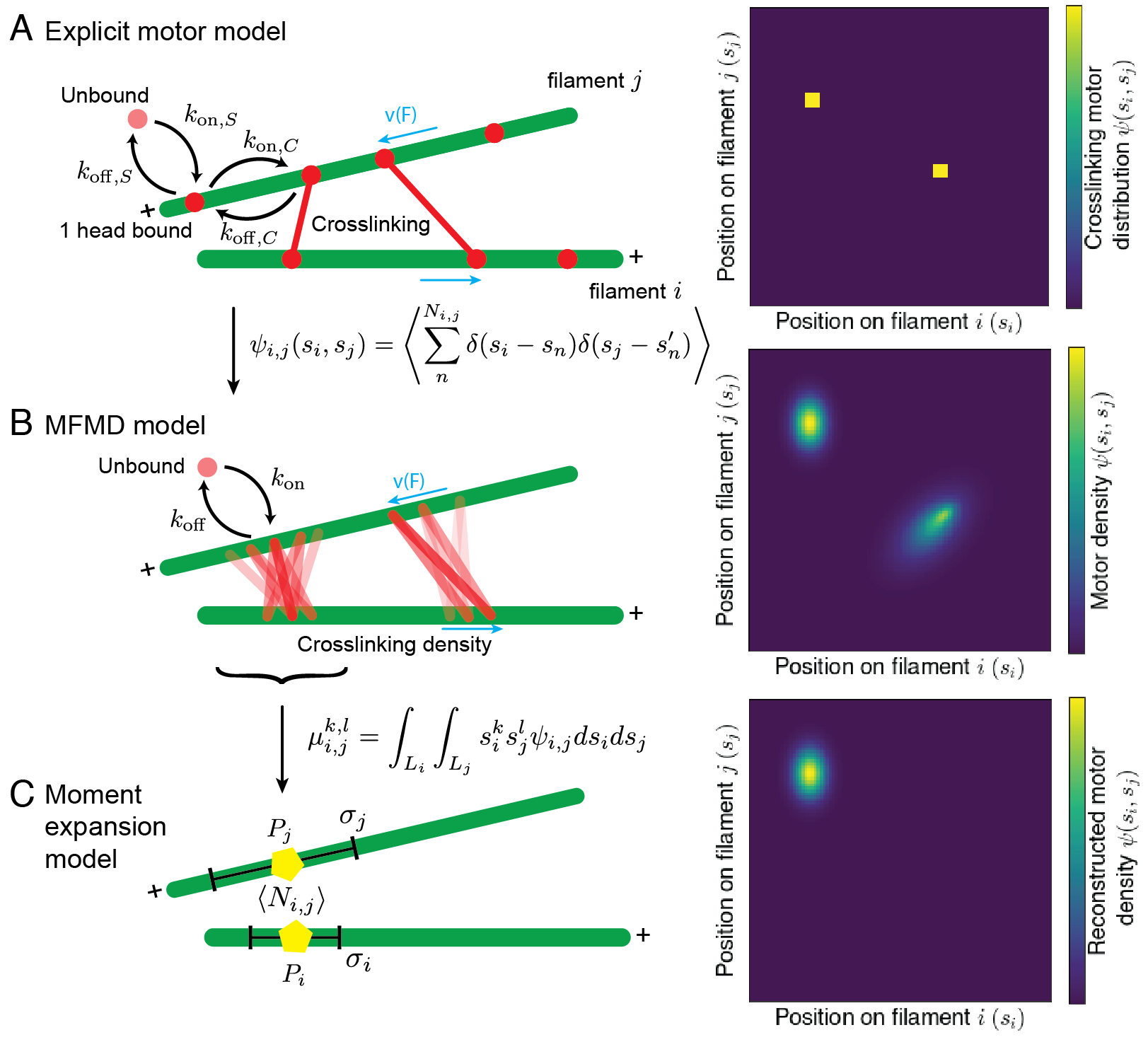}
  \caption{\footnotesize{Comparison of motor representations in three
      hierarchical models with schematics on the left and 2D motor
      distributions on the right.  {\bf A} Explicit motor model with two-step
      binding kinetics. Unbound motors~(light red circle) bind one head~(red
      circle) to filaments and then crosslink~(two red circles connected by red
      line).  {\bf B} Mean-field motor density model with motor
      distribution~(translucent red bars). Average motor distribution moments
      $\mu_{i,j}^{k,l}$ with respect to powers of bound crosslink positions
      $s_i$ and $s_j$. Moments are related to bound motor number
      $N_{i,j}$~(pentagon color), mean motor head position $P_i$~(pentagon
      position), and standard deviation $\sigma_i$~(black lines). (right) 2D
      plot of reconstructed motor density using bivariate Gaussian
      approximation.  For clarity, only moments derived from left-most
      crosslinking density distribution in ({\bf B}) are used to reconstruct 2D
      motor distribution in ({\bf C}.}}
  \label{fig:2_model_schematics}
\end{figure}

\section{Explicit motor model}%
\label{sec:molecular_model}
In the explicit motor model individual bound motors are modeled, allowing
fluctuations in bound motor number and binding kinetics that recover the
correct equilibrium distribution of crosslinking proteins in the limit of no
motor walking (Figure~\ref{fig:2_model_schematics}A)~\cite{gao15, gao15a,
  blackwell17a, rincon17, lamson19, edelmaier20}.

\subsection{Binding kinetics and stepping}%
\label{sub:binding_kinetics}
A motor diffuses in solution until one of its heads bind to a filament; we
model this by an infinite reservoir of unbound motors with a uniform and
constant concentration $c_o$. Filaments have a linear binding site density
$\epsilon$, and the binding site has an association constant $K_a$  (units of
$\mu$M$^{-1}$). First motor head binding has rate 
\begin{equation}
  \label{eq:KMC01imp}
  \kon[,S] = K_a c_o \epsilon L_{tot} \ko[,S],
\end{equation}
where  $L_{tot} = \sum_i L_i$ is the total length of filaments and
$\ko[,S]$ is the bare (force-independent) unbinding rate for singly bound heads.
All binding locations have equal binding probability. Singly bound motors unbind at 
 rate $ k_{{\rm off},S} = \ko[,S]$.

\begin{figure}[htpb]
  \centering
  \includegraphics[width=0.8\linewidth]{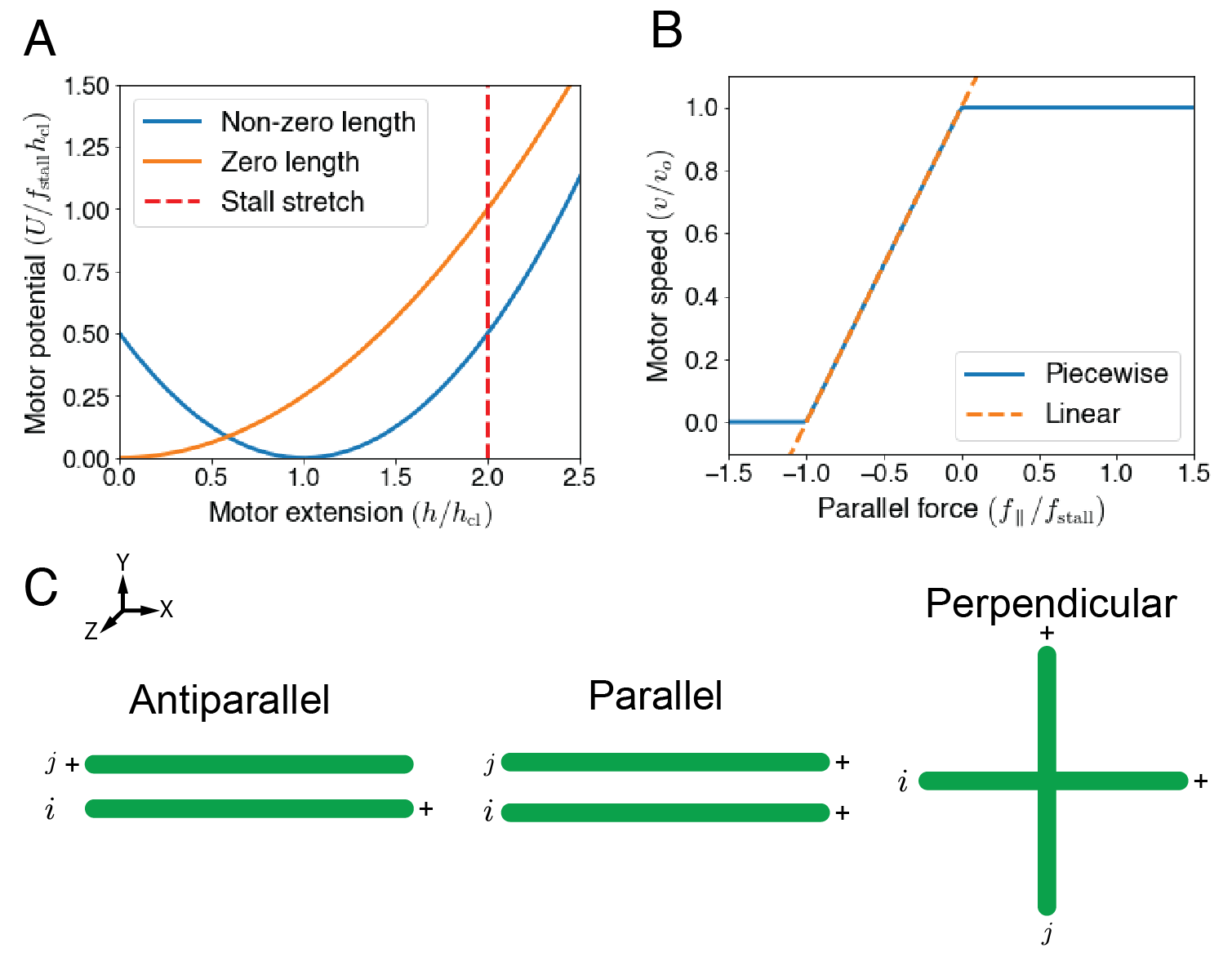}
  \caption{\footnotesize{Choice for motor tether potential, force-velocity
      relation, and filament initial configurations. {\bf A} Plot of
      the normalized potential energy in motor tether as a function of motor
      extension~(blue) and equivalent zero-length tether potential~(orange
      line).  Both potentials have identical slope at the distance
      $\Fs/\kcl$~(red dashed line) where motors stall. {\bf B} Plot of
      normalized motor speed as a function of force~(blue) and its linear
      approximation~(dashed orange). {\bf C} Chosen initial configurations of
      pairs of 1$\mu m$ filaments. Filament centers are separated by $D_{\rm
        fil}$ = 25nm perpendicular to both filament orientation vectors.}}%
  \label{fig:2-5_param_init_config}
\end{figure}

A motor with one head bound crosslink to another filament, which may stretch or
compress its tether. This makes crosslinking kinetics force dependent; our
models satisfy detailed balance in binding, so we recover the thermal
equilibrium Boltzmann distribution in the limit of passive crosslinkers. Motor
motion shifts the crosslinking distribution away from equilibrium.
Motor unbinding rate can depend on the force applied to bound
heads~\cite{Klumpp2005, Muller2008, Kunwar2011, Bouzat2016, Guo2019,
  Arpag2019}. Previous work shows how this force dependence can be included
while maintaining detailed balance in binding~\cite{grill05, blackwell17,
  edelmaier20}.  For simplicity, here we include the force dependence in the
binding rate only and discuss possible implications below.  With one head bound
to filament $i$ at position $s_i$, the free motor head binds to filament $j$ at
position $s_j$ with a probability proportional to a Boltzmann factor of binding
energy
\begin{equation}
  \label{eq:boltz}
  P_{S \to C}\propto \exp (-\beta \Ucl)
\end{equation}
with $\beta= (\kb T)^{-1}$ (Figure~\ref{fig:2-5_param_init_config}A). Here $S
\to C$ denotes the motor's transition from  a single head bound ($S$) to
crosslinking ($C$). The total binding rate is computed by integration over all
binding positions on filament $j$
\begin{equation}
  \label{eq:KMC_SD_Eindep}
  \kon[,C] = \frac{\epsilon \Ke \ko[,C]}{\Vbd} \int_{L_j} e^{-\beta \Ucl} ds_j,
\end{equation}
where  $\ko[,C]$ is the bare (force-independent) unbinding rate for a
crosslinking motor, $\Ke$ is the crosslinking  association constant. The
unbound motor head explores a volume $\Vbd$ centered about the bound head,
computed as  the integral of the unbound head's position weighted by the
Boltzmann factor
\begin{equation}
  \label{eq:double_bind_volume_gen}
  \Vbd = \int e^{-\beta \Ucl} dr^3 = 4\pi \int_0^{R_{{\rm cut},C}} e^{-\beta \Ucl} r^2dr.
\end{equation}
Beyond the cutoff radius $R_{{\rm cut},C}$ the integrand becomes small,
enabling the use of a lookup table
(Appendix~\ref{sec:creating_lookup_table_for_kmc_rm_s_to_d_}). The probability
distribution of binding position depends on the Boltzmann factor. We recover
the proper binding distribution through inverse transformation sampling of
equation (\ref{eq:KMC_SD_Eindep})
(Appendix~\ref{sec:reverse_lookup_algorithm}).

As discussed above, the unbinding rate of a single head of a motor crosslinking
two filaments is assumed to be force-independent,
\begin{equation}
  \label{eq:KMC_DS_Eindep}
  \koff[,C] = \ko[,C].
\end{equation}
Force-dependent unbinding affects the density of motor proteins
most when stretched\cite{Kunwar2011}; larger motor stretch occurs when external
force is applied against the force generated by motors. Therefore
sliding filaments slowed only by drag, like those in active nematics, will be
less affected by force-dependent unbinding than stationary filaments or jammed
filaments like microtubules in mitotic spindles.  We can include
force-dependent unbinding in the explicit motor and MFMD model but not in the
moment expansion model (Section~\ref{sec:moment_expansion_approximation}).
We chose the time step small enough that individual motors undergo only one
transition per time step
(Appendix~\ref{sec:determining_proper_time_step_for_binding}). 

The motor force-velocity relation is 
\begin{equation}
  \label{eq:vij}
  v_{i,j} = v(\uhat_i \cdot  \bff_{j,i} ) =
  \begin{cases}
    v_o, &  0 < \uhat_i \cdot  \bff_{j,i}  \\
    v_o \left( 1+ \frac{\uhat_i \cdot  \bff_{j,i} }{\Fs} \right) ,
  &         -\Fs < \uhat_i \cdot  \bff_{j,i} < 0 \\
  0, &       \uhat_i \cdot \bff_{j,i } < -\Fs, \\
  \end{cases}
\end{equation}
where $\Fs$ is the motor stall force (Figure~\ref{fig:2-5_param_init_config}B).

\subsection{Distribution of explicitly modeled motors}%
\label{sub:distribution_of_explicitly_modeled_motors}
The bound motor
distribution is
\begin{equation}
  \label{eq:explicit_psi}
  \psij(s_i, s_j, t) = \sum_{n=1}^{N_{i,j}(t)} \delta(s_i -
  s_n(t))\delta(s_j - s'_n(t)),
\end{equation}
where $\delta(s_i)$ is the Dirac delta function and $N_{i,j}$ is the total
number of motors crosslinking filaments $i$ and $j$. Here $s_n$ and
$s_n'$ are the attached positions of the heads of the $n$th crosslinking motor.
Motors with one head bound to filament $i$ have a distribution
\begin{equation}
  \label{eq:explicit_chi}
  \chi_i(s_i, t) = \sum_{n=1}^{N_i(t)} \delta(s_i - s_n(t)),
\end{equation}
where $N_i$ is the number of one-head bound motors on filament $i$. Only
motors crosslinking exert forces between filament pairs, but $\chi_i$ and
$\chi_j$ are needed to calculate the evolution of $\psij$.  

\section{Mean-field motor density model}%
\label{sec:fokker_planck_equation_for_motors}
Under typical experimental conditions, there can be tens to thousands of
crosslinking motors between a filament pair.  Motor force and torque
  fluctuations occur because of stochastic motor binding and
  unbinding. As the number of motors increases, the standard
  deviation relative to the mean decreases as $1/\sqrt{N}$. For
  our explicit motor model,  antiparallel filaments with an average of 14
  motors bound show a standard deviation in bound motor number of 27\% of the
  mean. This shows that the fluctuations are quite significant for order 10
  motors. The $1/\sqrt{N}$ scaling predicts that for an average of 1000 motors,
  the standard deviation would be only 3.2\% of the mean. The force and torque
  scale similarly.  
Therefore, for large motor number, we may use the average motor distribution to 
derive a mean-field motor density (MFMD) to accurately describe force and
torque on filaments by motors.  We can then evolve the MFMD instead of explicit
motors (Figure~\ref{fig:2_model_schematics}B). We previously showed that the
average steady-state density of crosslinking motors between stationary parallel
filaments agreed well with a solution to a multi-dimensional Fokker-Planck
equation (FPE)~\cite{blackwell16}. Here, we expand this approach to model
crosslinking motor density between filaments in three dimensions, allow
filament motion, and study time-dependent behavior of coupled systems of motors
and filaments.

For a one-step binding model, the MFMD evolves according to
\begin{equation}
  \label{eq:fpe}
  \pder[\psij(s_i, s_j, t)]{t} =
  - \pder[(v_{i,j}\psij)]{s_i} - \pder[(v_{j,i}\psij)]{s_j} + \kon - \koff\psij,
\end{equation}
with motor velocity $v_{i,j}$, motor crosslinking rate $\kon$, and unbinding
rate $\koff$.  To satisfy detailed balance in binding, we use the rates
$\kon=2\ko c e^{-\beta \Ucl(s_i, s_j)}$ and $\koff = 2\ko$, with the effective
concentration $c$ (units nm$^{-2}$)~\cite{blackwell16}. 
The factors of two
occur because there are two ways a motor can crosslink. To numerically solve
the hyperbolic equation (\ref{eq:fpe}), we use a first-order accurate upwind
difference method (Appendix~\ref{sec:numerical_solutions}).

The mean-field motor density model differs from the explicit model in that
motors with one head bound are not modeled explicitly.  To properly compare the
different binding models, we establish a mapping of parameters between these
two models (Appendix~\ref{sec:conversion_of_binding_parameters}), which gives 
\begin{equation} 
  \label{eq:concentration_compare} 
  c = \frac{\epsilon^2 K_a \Ke}{\Vbd}c_o.
\end{equation}
Some model parameters are difficult to measure directly. For example, the
association constant $\Ke$ may differ from $K_a$ if proteins change their
molecular conformation when bound.  We discuss an approach to estimate such parameters
in Appendix~\ref{sub:calculating_binding_parameters_from_experiments}.

\subsection{Steady-state solution for MFMD on antiparallel filaments}%
\label{sub:steady_state_of_motors}
If filaments move slowly compared to the timescale of motor rearrangement, then
a quasi-steady state approximation can be used.  In the quasi-steady limit, the
force and torque on filaments are computed from the steady-state
MFMD~\cite{lamson19}. The quasi-steady  approximation is computationally
efficient compared to numerical integration of the time-dependent PDE. A
steady-state solution also provides a convenient route to compare our model
implementations. 

At steady state, equation~(\ref{eq:fpe}) becomes
\begin{equation}
  \label{eq:fpss}
  \psij\pder[v_{i,j}]{s_i} + v_{i,j} \pder[\psij]{s_i} +
  \psij\pder[v_{j,i}]{s_j} + v_{j,i} \pder[\psij]{s_j} +
  2\ko \psij = 2\ko c e^{-\beta U(s_i, s_j)}.
\end{equation}
Here we choose functional forms of $\Ucl$ and
$v_{i,j}$ consistent with previous
models~\cite{blackwell16,popov16,rincon17,freedman17,lamson19}. Motors have a
potential energy
$
  \Ucl  = \frac{\kcl}{2}\left( h_{i,j} - \hcl\right)^2
$
determined by the tether spring constant $\kcl$ and tether length
$\hcl$~(Figure~\ref{fig:2-5_param_init_config}A), which implies a  motor
crosslinking filaments $i$ and $j$  exerts a force on $\bff_{i,j} = -\kcl
\left( 1 - \frac{\hcl}{h_{i,j}} \right) \bh_{i,j} $ on filament $j$. The
force-velocity relation of a motor head attached to filament $i$ while the
other head is bound to $j$ follows equation (\ref{eq:vij}).  Here, we assume
motors that reach filament ends walk off, i.e.,\ no end pausing.  

A semi-analytic steady-state solution can be derived for antiparallel filaments
when motor tethers have zero length ($\hcl = 0$) because the FPE is symmetric
under the transformation $i \to j$.  For zero-tether-length motors to mimic
their non-zero-length counterparts, we modify the zero-length motor's spring
constant so both types of motors stall at the same extension $h_{i,j} = \hs$.
This implies $\kcl' \hs = \kcl ( \hs - \hcl) = \Fs$ with the solution  
\begin{equation}
  \label{eq:kcl_stretch_equiv} 
  \kcl' = \frac{\kcl\Fs}{\Fs + \kcl \hcl},
\end{equation} 
where $\hs = \Fs/\kcl'$.  Note this choice changes the binding dynamics,
because the potential energy is now larger for larger motor
extension~(Figure~\ref{fig:2-5_param_init_config}A). 

To find the steady-state solution, note that $\ru, \ru[j][i] = 0$ and $\cb =
-1$ for antiparallel filaments with centers aligned.  Therefore, $\hij =
\sqrt{\rsqr + (s_i + s_j)^2}$ and $\uhat_i\cdot\bff_{j,i} =
\uhat_j\cdot\bff_{i,j} = -\kcl'(s_i+s_j)$. Since $U_{i,j}$, $v_{i,j}$, and
$v_{j,i}$ depend exclusively on the sum of $s_i$ and $s_j$, we make the change
of variables $\xi=s_i + s_j$ in equation (\ref{eq:fpss}) to find
\begin{equation}
  \label{eq:fpss_apara}
  \left( v_{i,j} + v_{j,i} \right) \pder[\psij]{\xi} +
  \left( \pder[v_{i,j}]{\xi} + \pder[v_{j,i}]{\xi} + 2\ko \right) \psij
  = 2\ko c e^{-\beta U(\xi)}.
\end{equation}
There are three regions of solution determined by the force-velocity relation
equation (\ref{eq:vij}): $\xi \leq 0$, $0 \leq \xi \leq \hs$, and $\hs < \xi$.
For $\xi \le 0$, $v_{i,j} = v_{j,i} = v_o$ and equation (\ref{eq:fpss_apara})
becomes
\begin{equation}
  \label{eq:fpss_apara_reg1}
  \pder[\psij]{\xi} +  \frac{\psij}{l_o} = \frac{c}{l_o} e^{-\beta U(\xi)},
\end{equation}
where $l_o = v_o/\ko$ is the motor run length. This is solved with an
integrating factor, giving
\begin{equation}
  \label{eq:fpss_apara_reg1_sol}
  \psij(\xi) = e^{ \frac{\xi - L}{l_o}}\psij(-L) + \frac{c}{l_o} e^{-\xi/l_o}\int_{-L}^{\xi}e^{ \frac{x}{l_o} - \frac{\beta \kcl'}{2} (\rsqr+x^2)}dx.
\end{equation}
Applying the boundary condition $\psij(-\frac{L}{2},-\frac{L}{2}) = \psij(-L) =
0$, we remove the last term in equation (\ref{eq:fpss_apara_reg1_sol}) and
re-write the Gaussian integral as
\begin{equation}
  \label{eq:fpss_apara_reg1_simp}
  \psij(\xi) = \frac{c}{l_o}\sqrt{\frac{\pi}{2\beta\kcl'}}
  \exp \left( \frac{1}{2\beta \kcl' l_o^2} - \frac{\beta \kcl'}{2}r^2 -\frac{\xi}{l_o} \right)
  \left[ \erf \left( \frac{\beta\kcl' l_o x - 1}{l_o\sqrt{2\beta\kcl'}}\right) \right]_{x=-L}^{x=\xi}.
\end{equation}

For $0 \leq \xi \leq \hs$, the velocity $v_{i,j} = v_{j,i} = 1
-\frac{\xi}{\hs}$. Equation (\ref{eq:fpss_apara}) becomes
\begin{equation}
  \label{eq:fpss_apara_reg2}
  \left( \hs - \xi \right) \pder[\psij]{\xi} + \left( \frac{\hs}{l_o} - 1 \right) \psij
  = \frac{\hs}{l_o}ce^{-\beta U(\xi)}.
\end{equation}
Solving with an integrating factor, we find
\begin{equation}
  \label{eq:fpss_apara_reg2_sol}
  \psij(\xi) = \psij(0)\left( \frac{\hs}{\hs+\xi} \right)^{1 - \frac{\hs}{l_o}} + \frac{\hs c}{l_o(\hs - \xi)^{1-\frac{\hs}{l_o}}}
              \int_{0}^{\xi}(\hs - x)^{-\frac{\hs}{l_o}}e^{- \frac{\beta \kcl'}{2} (\rsqr+x^2)}dx.
\end{equation}
We match the solution for $\psij(0)$ to equation~(\ref{eq:fpss_apara_reg1_sol})
to enforce continuity.  The exponential term in equation
(\ref{eq:fpss_apara_reg2_sol}) can be approximated by a series expansion or
integrated numerically. Here we use numerical integration.

For $\xi > \hs$, the velocity and velocity derivatives are zero, so
\begin{equation}
  \label{eq:fpss_apara_reg3_sol} \psij(\xi) = c e^{- \frac{\beta \kcl'}{2}
    (\rsqr+\xi^2)}.  
\end{equation} 
Since the motor velocity is zero at $\xi=\hs$, motors do not walk from $\xi <
\hs$ to $\xi>\hs$. A non-zero MFMD exists for $\xi>\hs$ only if motors bind at
these lengths. This appears as an integrable discontinuity at $\xi = \hs$.

\section{MFMD moment expansion}%
\label{sec:moment_expansion_approximation}
A series expansion or reduced representation of a continuous distribution can
lower the computational cost of solving a system's time evolution~\cite{wang11,
  mathijssen16, gao17}. Here, we use low-order moments of the MFMD  to
calculate motor number, mean and standard deviations of motor head
distribution, and filament motion.

The moments of $\psij$ are
\begin{equation}
  \label{eq:psi_moments}
  \mu_{i,j}^{k,l}(t) = \int_{L_{i}} \int_{L_{j}} s_i^k s_j^l \psij ds_i ds_j,
\end{equation}
where $k,l$ are non-negative integers. The moments are symmetric under exchange
of both filaments and powers so that $\mu_{i,j}^{k,l} = \mu_{j,i}^{l,k}$. The
zeroth moment $\mu_{i,j}^{0,0} = N_{i,j}$ is the total number of motors bound
to the two filaments, and the first moments $\mu_{i,j}^{1,0}, \mu_{i,j}^{0,1}$
are proportional to the mean motor head position along each filament
$P_i=\frac{\mu_{i,j}^{1,0}}{N_{i,j}}, P_j=\frac{\mu_{i,j}^{0,1}}{N_{i,j}}$.
The first two second moments determine the standard deviation of motor head
density
\begin{equation}
  \label{eq:sigma_i}
  \sigma_i = \sqrt{\frac{\mu_{i,j}^{2,0}}{N_{i,j}}- P_i^2}.
\end{equation}
The symmetric second moment term $\mu_{i,j}^{1,1}$ determines the covariance of
motor head position
\begin{equation}
  \label{eq:Vij}
  V_{i,j} = \frac{\mu_{i,j}^{1,1}}{N_{i,j}}-P_i P_j.
\end{equation}
The positional means, standard deviations, and covariance are used to
reconstruct an approximate MFMD for visualization using a bivariate normal
distribution (Figure~\ref{fig:2_model_schematics}C, Videos~1-6).

Using the approximation of zero-length tethers as in
section~(\ref{sub:steady_state_of_motors}) above, $\bff_{i,j}$ is a linear
function of $s_i$ and $s_j$.  In this case, filament motion can be computed
from low-order moments using equations (\ref{eq:F_mean}) and (\ref{eq:T_mean}):
\begin{align}
  \label{eq:force_moment}
  \begin{split}
  \bF_{i,j} &= -\kcl \int_{L_{i}} \int_{L_{j}} \left( \Dr + s_j\uhat_j -s_i\uhat_i \right)\psij ds_i ds_j\\
  &= -\kcl \left( \mu_{i,j}^{0,0}\Dr  + \mu_{i,j}^{0,1}\uhat_j - \mu_{i,j}^{1,0}\uhat_i \right)
  \end{split}
\end{align}
and
\begin{align}
  \label{eq:torque_moment}
  \begin{split}
    \bT_{i,j} &= -\kcl \int_{L_{i}} \int_{L_{j}} s_j \hat{u}_j \times \left( \Dr + s_j\uhat_j -s_i\uhat_i  \right) \psij ds_i ds_j\\
      &= -\kcl \uhat_j \times \left( \mu_{i,j}^{0,1}\Dr  - \mu_{i,j}^{1,1}\uhat_i \right)
  \end{split}
\end{align}
Substituting equations (\ref{eq:force_moment}) and (\ref{eq:torque_moment})
into equation (\ref{eq:dot_r_i}) and (\ref{eq:dot_uhat_i}) show that only
moments up to second order are needed to compute filament motion from
crosslinking motors. Thus, motor and filament evolution can be written as a
system of ODEs that depend on the dynamical evolution of the moments. This
dynamical evolution is computed by taking the time derivative of equation
(\ref{eq:psi_moments}) and substituting in the FPE (\ref{eq:fpe}) 
\begin{equation} 
  \label{eq:dot_mukl_gen}
  \pder[\mu_{i,j}^{k,l}]{t} = \int_{L_{i}} \int_{L_{j}} s_i^k s_j^l \pder[\psij]{t} ds_i ds_j.
\end{equation}
However, this coupled system of equations for the moment time evolution  does
not close.  Because the piecewise motor force-velocity relation is not linear,
moments depend on higher-order moments recursively.  
Also, filament ends
introduce boundary terms that prevent closure. Despite this, in certain limits
a truncated moment expansion shows good agreement with the explicit and MFMD
models.  

We first introduce a linear approximation to the force-velocity relation
(Figure~\ref{fig:2-5_param_init_config}B)
\begin{eqnarray}
  \label{eq:v_linear}
  v_{i,j} \approx v_o \left( 1+\frac{\uhat_i \cdot \bff_{j,i}}{\Fs} \right) =
  v_o \left( 1+ \frac{\kcl}{\Fs}\left( \ru + \cb s_j - s_i \right) \right).
\end{eqnarray}
This approximation is valid for $\hs \gg \sqrt{1/\kcl \beta}$, in which case
motors do not bind beyond their stall stretch. We also require that $v_o \gg
2\ko\sqrt{1/\kcl \beta}$, ensuring that motors pulled towards the plus ends
with $\uhat_i\cdot\bff_{i,j} > 0$ move quickly into a regime $-\Fs <
\uhat_i\cdot\bff_{i,j} < 0$, where the linear and piecewise force-velocity
functions agree.

We substitute the linearized force-velocity function from equation
(\ref{eq:v_linear}) into the MFMD equation (\ref{eq:fpe}) to obtain
\begin{align}
  \label{eq:fpe_lin_v}
  \begin{split}
    \pder[\psij]{t} =  2\ko c e^{-\beta \Ucl} + \left( 2\kappa - 2\ko \right)\psij
              &- \left( v_o +\kappa \left( \ru + \cb s_j - s_i \right) \right)\pder[\psij]{s_i} \\
              &- \left( v_o +\kappa \left( \ru[j][i] + \cb s_i - s_j  \right) \right)\pder[\psij]{s_j},
  \end{split}
\end{align}
where $\kappa = v_o \kcl/\Fs$ is the rate at which motors reach their stall
force. Integrating equation (\ref{eq:fpe_lin_v}) directly returns the zeroth
moment equation
\begin{align}
  \label{eq:zero_moment}
  \begin{split}
  \pder[\mu_{i,j}^{0,0}]{t} = 2\ko \q^{0,0} -2\ko \mu_{i,j}^{0,0}
                              &+ {\left[ \left( -v_o - \kappa \ru[j][i] + \kappa s_i\right) B_j^0
                                            - \kappa \cb B_j^1 \right]}_{\del L_{i}} \\
                                  &+ {\left[ \left( -v_o - \kappa \ru + \kappa s_j\right) B_i^0
                                            - \kappa \cb B_i^1 \right]}_{\del L_{j}},
  \end{split}
\end{align}
where we have defined $\q^{k,l}=\int_{L_{i}}\int_{L_{j}}s_i^k s_j^l e^{-\beta \Ucl}ds_i ds_j$ and
\begin{equation}
  \label{eq:bjl}
  B_j^l(s_i) = \int_{L_{j}}s_j^l \psij ds_j
\end{equation}
with $\q^{k,l}$ representing source terms. Here $B_j^l(s_i)$ is a moment of the
MFMD integrated over $s_j$ that is a function of $s_i$, but in practice $B_j^l$
only appears in the equations evaluated at filament endpoints, and so captures
behavior of the motor density at filament ends. Therefore we refer to the
$B_j^l(s_i)$ as boundary terms. To show this, we define the notation
$[A(s_i)]_{\del L_i} = A(L_i/2) - A(-L_i/2)$.

The general moment evolution obtained by integrating equation
(\ref{eq:dot_mukl_gen}) with equation (\ref{eq:fpe_lin_v}) is
\begin{align}
  \label{eq:psi_moments_dot}
  \begin{split}
    \pder[\mu_{i,j}^{k,l}]{t} = 2\ko \q^{k,l} & + k(v_o + \kappa \ru)\mu_{i,j}^{k-1,l}
                            + l(v_o + \kappa \ru[j][i])\mu_{i,j}^{k,l-1} \\
          & - \left( 2\ko +(k+l)\kappa \right)\mu_{i,j}^{k,l}
            + \kappa \cb \left( k\mu_{i,j}^{k-1,l+1} + l\mu_{i,j}^{k+1,l-1} \right)\\
          & + \left[ \left( \kappa s_i^{k+1} - \kappa \ru s_i^{k} - v_o s_i^{k} \right) B_j^l
            - \kappa \cb s_i^k B_j^{l+1} \right]_{\del L_{i}} \\
          & + \left[ \left( \kappa s_j^{l+1} - \kappa \ru[j][i] s_j^l - v_o s_j^l \right) B_i^k -
            \kappa \cb s_j^l B_i^{k+1} \right]_{\del L_{j}}.
  \end{split}
\end{align}
The boundary terms in square brackets contain moments and $B_j^l$ an order
higher than $\del\mu^{k,l}_{i,j}/\del t$.  In
Appendix~\ref{sub:boundary_value_calculations}, we write the analogous time
evolution for the $B_j^l$, and show that it does not close. Therefore the
moment evolution equations do not close.  

To close the system of equations, we  set the boundary terms to zero.
Physically, this means we neglect motor unbinding from filament plus ends. If
motors pause at plus ends, this approximation will lead to significant error.
However, if motor unbinding is relatively rapid (including at filament plus
ends), this is a good approximation. To explore the impact of not including
these boundary terms, below we quantify the discrepancy between this model and
the explicit motor and MFMD models. Neglecting boundary terms truncates the
system of equations at second order, because only terms up to second order are
needed to calculate force and torque on filaments. 

We evolve equations
(\ref{eq:dot_r_i},~\ref{eq:dot_uhat_i},~\ref{eq:psi_moments_dot}) using
\verb|solver_ivp|  in the \verb|scipy.integrate| library~\cite{scipy20}. This code uses the
LSODA integrator, an Adams/BDF integration method that automatically detects
stiffness, from the Fortran \verb|ODEPACK| library~\cite{hindmarsh83}.  The
source terms $\q^{k,l}$ are analytically integrated in one dimension and then
numerically integrated using the quad method also from \verb|scipy.integrate|
(Appendix~\ref{sec:speed_up_of_gaussian_integrals_in_moment_expansion}).

\section{Results}%
\label{sec:results}

\input{params_units_tab.tex}

To test the degree of agreement between explicit motor and mean-field models,
we first selected parameters based on microtubules and kinesin-5 motor proteins
because they are relatively well-studied cytoskeletal proteins~\cite{kashina96,
  kawaguchi01, valentine06, cochran15} (Table~\ref{tab:units}).  We studied
three characteristic sets of initial filament pair position and orientation:
antiparallel, parallel, and perpendicular
(Figure~\ref{fig:2-5_param_init_config}, Video~1-3), and compared both
stationary and moving filaments. 
We choose an initial condition with no motors bound to filaments, in order to
observe the effects of time evolution of the motor density.  For stationary
filaments, we found good
agreement for all three models. For moving filaments, we found qualitative
agreement but fluctuations in motor dynamics and different end boundary
conditions contributed to quantitative differences in filament motion.  We
measured the computational cost for stationary antiparallel filaments and found
that the moment-expansion model can give a dramatic improvement in performance.

\subsection{Stationary filament pairs}%
\label{sub:stationary_filament_pairs}

\begin{figure}[htpb]
  \centering
  \includegraphics[width=.7\linewidth]{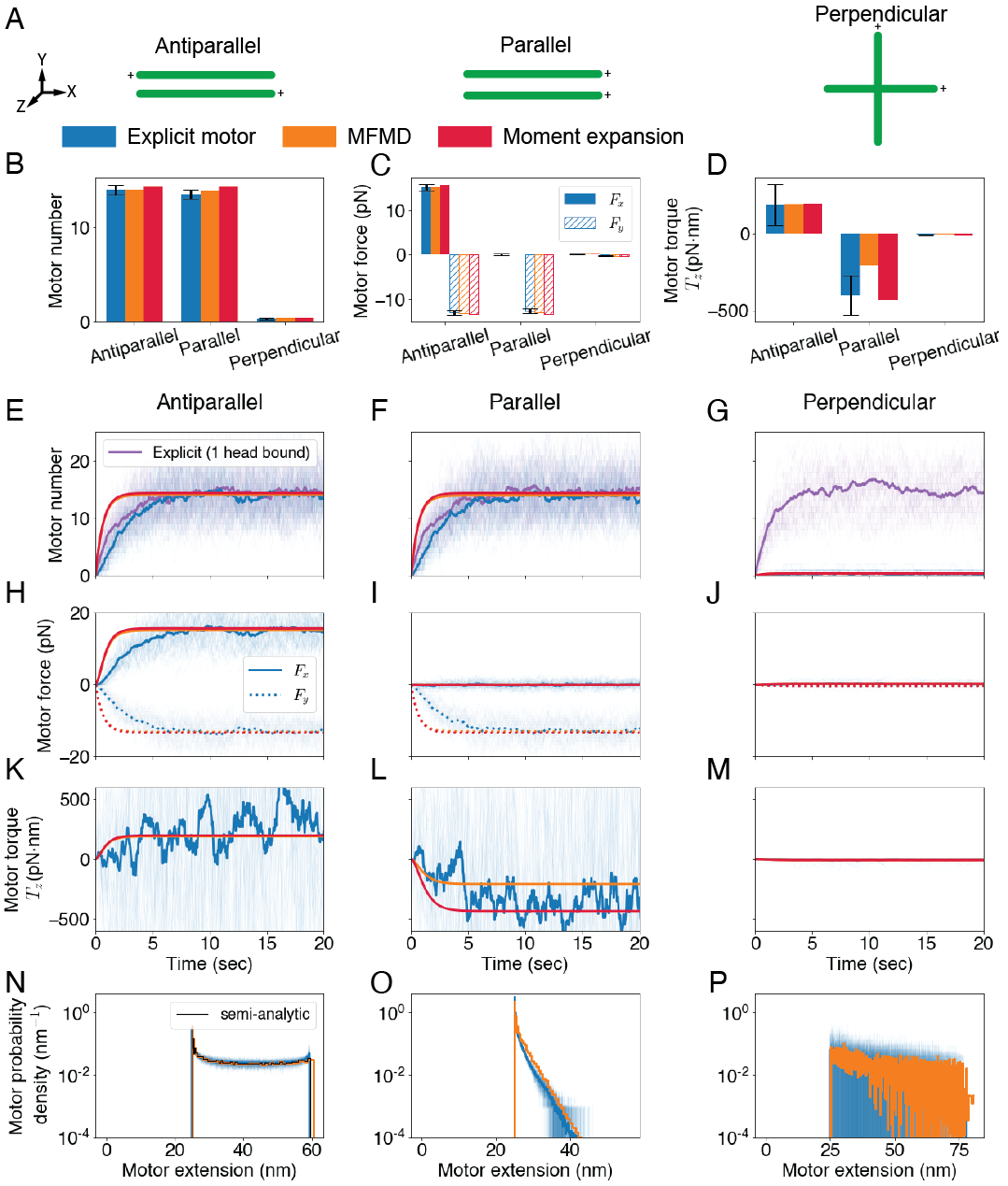}
  \caption{\footnotesize{Comparison of model results for three different
      stationary filament configurations. {\bf A} Schematic of the three
      different filament configurations and legend for following plots.  {\bf
        B} Plot of total crosslink motor numbers at steady state.  {\bf C} Plot
      of steady-state motor force components from filament $i$ on filament $j$.
      {\bf D} Bar graph of steady-state torque in the $\hat{z}$-direction from
      filament $i$ on filament $j$.  Explicit motor model error bars in (B-D)
      indicate the Standard Error of the Mean (SEM) of the last 30 seconds of
      40 second long simulations~(n=48). {\bf E-G} Bound motor number versus
      time. Purple and blue solid lines are the average of 48 individual
      explicit motor simulations~(translucent lines) for one head bound and
      crosslinking motors.  {\bf H-J} Motor force in the
      $\hat{x}$-direction~(solid lines) and $\hat{y}$-direction~(dotted lines).
      Individual explicit motor runs are represented as blue for both
      directions.  {\bf K-M} Motor torque in the $\hat{z}$-direction from
      filament $i$ on $j$. Full explicit motor model range not shown to better
      see average. {\bf N-P} Steady-state motor probability density as a
      function of motor extension for semi-analytic~(black), explicit motor,
      and MFMD models. Motor minimum extension is set by the separation of
      filaments at closest point of approach, 25 nm.}}
  \label{fig:3_stat_compare}
\end{figure}

When filaments are held stationary, motor density reaches or fluctuates around
a steady-state solution (Figure~\ref{fig:3_stat_compare}). 
To compare with the
mean-field models, we averaged 48 realizations of each explicit motor
simulation; the results agreed within error with the mean-field models
(Figure~\ref{fig:3_stat_compare}B-D). This agreement between models
demonstrates that the mean-field models capture the average behavior of our
explicit model. 

Beyond the steady state, we characterize the evolution of motor number, force,
and torque (Figure~\ref{fig:3_stat_compare}E-M). In all configurations, the
crosslinking motor number in the explicit motor model lags that of the MFMD and
moment expansion models (Figure~\ref{fig:3_stat_compare}E-G). The crosslinking
rate in the two-step binding algorithm depends on the density of motors with
one head bound, resulting in a slower  approach to steady state.  

For antiparallel filaments, force generation increases with crosslinking motor
number (Figure~\ref{fig:3_stat_compare}H, Video~1) because motors walk in
opposite directions, causing the motor tether to stretch and generate force. If
free to move, these antiparallel filaments would slide. No average sliding
would occur for parallel filaments, and the small number of crosslinking
proteins for perpendicular filaments results in small relative force
(Figure~\ref{fig:3_stat_compare}I, J).  The average explicit motor motor torque
in the $\hat{z}$-direction shows significant fluctuations about the mean
(Figure~\ref{fig:3_stat_compare}K-M). Because motor torque increases for motors
farther from the filament centers, the torque fluctuations increase with
filament length. 

We  compared the steady-state distribution of motor extension for both explicit
motor and MFMD models (Figure~\ref{fig:3_stat_compare}N-P). (Note that the
moment expansion loses this information in coarse-graining.) The distribution
of motors crosslinking antiparallel filaments has two peaks
(Figure~\ref{fig:3_stat_compare}N).  The larger peak represents the most
probable binding distance $\Delta y$, and the second peak corresponds to motors
near their stall extension $h=\sqrt{\Delta y^2 + \hs^2}$.  The shape of the
distribution results from motor kinetics, walking, and stalling.  Motors on
parallel filaments  show a  peak at $\Delta y$
(Figure~\ref{fig:3_stat_compare}O, Video~2), but no second peak because the
motor heads walk in the same direction with similar speed. For motors
crosslinking perpendicular filaments, the extension distribution is singly
peaked and  broader than for parallel filaments
(Figure~\ref{fig:3_stat_compare}P, Video~3). This occurs because the parallel
force component on perpendicular filaments increases more gradually as the
motors extend, causing a more gradual decrease in  motor speed. This broad
distribution indicates a larger average force per motor for perpendicular
filaments compared to aligned filaments.

\subsection{Dynamical evolution of filament pairs}
\label{sub:constrained-motion_filament_pairs}

\begin{figure}[htpb]
  \centering
  \includegraphics[width=.8\linewidth]{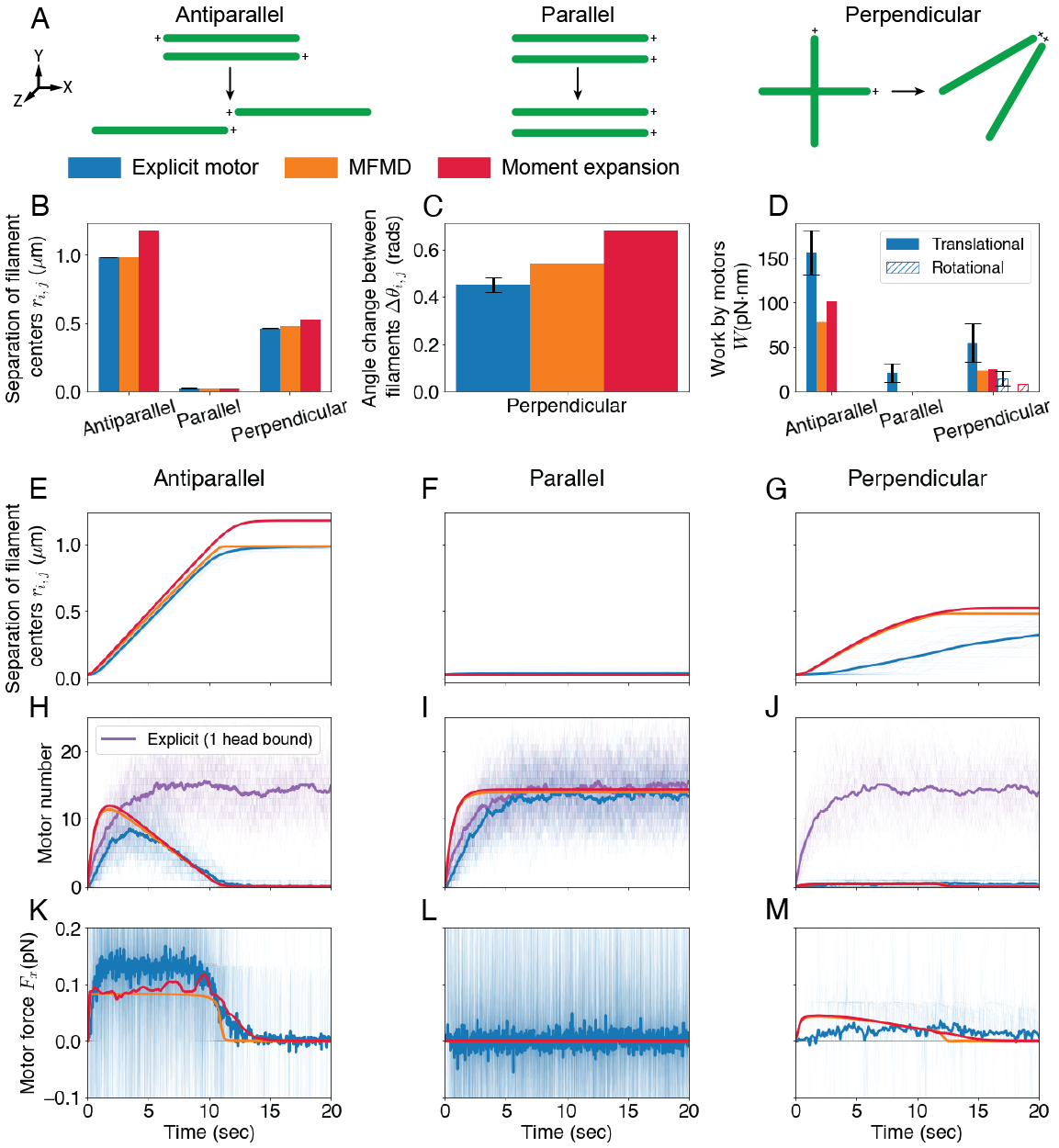}
  \caption{\footnotesize Comparison of model results for three different
    initial filament configurations evolved with constrained motion.  {\bf A}
    Schematic of initial and final filament configurations.  {\bf B} Plot of
    final filament center separations. {\bf C} Plot of change in angle between
    filaments starting in a perpendicular configuration. Data shown is final
    configuration after 100 seconds for the explicit motor model and 20 seconds
    for MFMD and moment expansion model. {\bf D} Plot of translational~(solid
    bars) and rotational~(hatch bars) work done by motors on filaments during
    simulation.  explicit motor model error bars in (B-D) indicate the SEM of
    simulation realizations~(n=48). {\bf E-G} Plots of filament centers
    separation as a function of time. {\bf H-J} Plots of motor number versus
    time.  Purple and blue solid lines are the average of 48 explicit motor
    simulations~(translucent lines) for one head bound and crosslinking motors.
    {\bf K-M} Plots of motor force in the $\hat{x}$-direction with individual
    explicit motor runs~(translucent blue lines) and average~(solid blue). Full
    explicit motor model range not shown to better see average.}%
  \label{fig:4_cstr_compare}
\end{figure}

Here we consider the same three filament starting configurations and allow
filament motion (Figure~\ref{fig:4_cstr_compare}, Videos~4-6). The final
filament position and orientation are comparable for the explicit motor and
MFMD models, while the moment expansion model overestimates the range of
filament translation and rotation (Figure~\ref{fig:4_cstr_compare}B, C; note
that filament rotation only occurs for the perpendicular initial
configuration). 

To compare motor activity between models over the whole simulation, we
calculated the total work done by motors. We numerically integrate both
filaments using the trapezoid rule~\cite{siyyam97},
\begin{equation}
  \label{eq:work_calc}
  W_{{\rm tot}} = W_{{\rm lin}} + W_{{\rm rot}}
  = \sum_{i \neq j} \int \bF_{i,j} \cdot d\br_j
  + \sum_{i \neq j} \int \bT_{i,j} \cdot d{\bm \theta}_{j},
\end{equation}
where $\theta_j$ is the angle the vector $\uhat_j$ rotates through over the
simulation. The infinitesimal vector $d{\bm \theta}_i= \hat{\theta}_i d\theta_i$
where 
\begin{equation}
  \label{eq:theta_i}
  \hat{\theta_i} = \frac{\uhat_i \times \dot{\uhat}_i}{|\uhat_i \times \dot{\uhat}_i|}.
\end{equation}
Total work computed for the mean-field models is within error of the explicit
motor model (Figure~\ref{fig:4_cstr_compare}D).  We note that the explicit
motor model produces greater total work because fluctuations in motor binding
cause fluctuations in sliding direction which generate larger work. Motors
generate rotational work only for initially perpendicular filaments, due to
the constraints. The magnitude of the rotational work is relatively small
because filaments rotate slowly (due to high rotational drag and low motor
torque), and this slower velocity produces less work in the overdamped limit. 

The crosslinking motor number depends on the filament overlap length, which
changes as filaments move (Figure~\ref{fig:4_cstr_compare}E-J).  The
crosslinking motor number in the explicit motor model lags the mean-field
models initially due to differences in binding, but becomes comparable after
the initial transient.  As antiparallel filaments slide apart, their overlap
decreases so fewer motors crosslink, while crosslinking motors continue to
unbind at a constant rate.  However, the overlap length has little effect on
the number of motors with one head bound (Figure~\ref{fig:4_cstr_compare}E, H).
The dynamics of motor number for parallel stationary and moving filaments are
nearly identical because there is negligible sliding.
(Figure~\ref{fig:4_cstr_compare}F, I). Moving perpendicular filaments maintain
a similar overlap length to  stationary perpendicular filaments, leading to an
approximately constant motor number, until the plus-ends move close together
(Figure~\ref{fig:4_cstr_compare}G, J). Then motors continue to bind but
immediately walk off, producing little force or torque.

The motor force between antiparallel filaments rapidly reaches a force plateau
which persists until the antiparallel overlap length is small enough that motor
binding is negligible (Figure~\ref{fig:4_cstr_compare}K).  The nearly constant
force implies that  motor extension decreases as the number of crosslinking
motors increases to give a constant sliding speed
(Figure~\ref{fig:4_cstr_compare}N, Video~4). This steady-state force is an
order of magnitude smaller than the stall force (Table~\ref{tab:units}).  The
moment expansion model shows a slower decrease in force as the overlap
approaches zero compared to the MFMD model (Figure~\ref{fig:4_cstr_compare}H).
This is a consequence of our neglect of boundary terms, which physically means
neglecting motor dissociation at  filament ends.  This unphysical slow force
decrease drives filaments beyond the zero overlap configuration to larger than
expected separation (Figure~\ref{fig:4_cstr_compare}B).

Parallel filaments remain with their centers aligned on average because
sampling the full distribution of motor crosslinking extension generates
restoring force for any fluctuations away from full overlap  (equations
\ref{eq:KMC_SD_Eindep},~\ref{eq:KMC_DS_Eindep}).  Neither the MFMD nor the
moment expansion models produce a net force, but in the explicit motor model
fluctuations in motor number and binding lead to force and position
fluctuations (Figure~\ref{fig:4_cstr_compare}F, I, L, Video~5).  For
perpendicular filaments, the small number of crosslinking motors results in
large force fluctuations in the explicit motor model
(Figure~\ref{fig:4_cstr_compare}J).  The mean-field models show a rapid
increase to half the maximum force of the antiparallel configuration followed
by a decrease  as the filaments align parallel
(Figure~\ref{fig:4_cstr_compare}K, M).  The lag caused by the two-step binding
model is more apparent here because the explicit lower motor number means
filaments move more slowly into the parallel configuration where binding is
favored (Video~6).

\subsection{Computational cost and accuracy}%
\label{sub:computational_performance_and_accuracy}

\begin{figure}[htpb]
  \centering
  \includegraphics[width=0.9\linewidth]{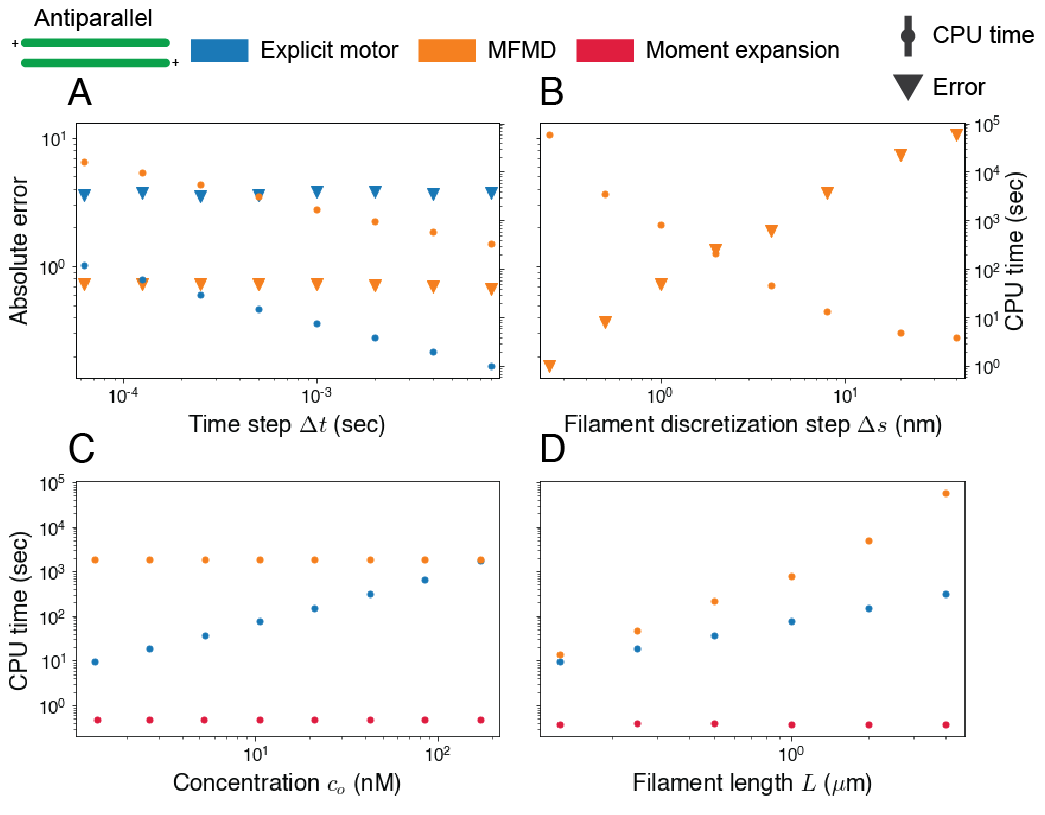}
  \caption{\footnotesize{Comparison of computational cost and accuracy of
      models for stationary filaments in an antiparallel configuration. Error
      compared against steady-state solution.  {\bf A} Plot of error (triangle)
      and CPU time (circle) vs time step $\Dt$ for explicit motor and MFMD
      models. Each for explicit motor model data point consists of 48 parameter
      set realizations. MFMD simulations were run 3 times to ensure consistency
      of time scaling. Standard error of the mean (SEM) of CPU time plotted but
      not visible.  {\bf B} Plot of error and CPU time vs $\Ds$ for MFMD model.
      Simulations were run 3 times to ensure consistency. SEM of CPU time
      plotted but not visible.  {\bf C, D} Plot of CPU time vs unbound motor
      concentration $c_o$ and filament length $L$ for the three models.
      Explicit motor simulation data points in C and D consist of 24 parameter set
      realizations while MFMD and moment expansion data points consist of 3
      runs for concentration and 4 runs for filament length. SEM of CPU time
      plotted but not visible.}}%
  \label{fig:5_performance}
\end{figure}

To compare the accuracy and computational cost of our models, we focus on
stationary antiparallel filaments because we can compare to the semi-analytic
solution. Antiparallel filaments are also the main configuration in which
motors generate extensile force, important for mitotic spindle assembly and
dynamics in active nematics.  We vary the time step $\Dt$ and MFMD grid spacing
$\Ds$ and compare the error with the semi-analytic solution. The
central-processing unit (CPU) time measures the computational cost as a
function of simulation parameters.

The solution error is the average magnitude of the deviation of the
steady-state numerical solution from $\psij$ of equations
(\ref{eq:fpss_apara_reg1_sol}),
(\ref{eq:fpss_apara_reg2_sol}), and (\ref{eq:fpss_apara_reg3_sol}), 
\begin{equation}
  \label{eq:error}
  {\rm Error} = \int_{L_i}\int_{L_j} \left|\overline{\psi}_{i,j} - \psij \right| ds_i ds_j
  \approx \sum_{m,n} \left| \overline{\psi}_{i,j}(m\Ds_i, n\Ds_j) -\psij(m\Ds_i, n\Ds_j) \right|\Ds_i \Ds_j,
\end{equation}
where $\overline{\psi}_{i,j}$ is either the average explicit motor distribution
(over 48 simulations) or the MFMD distribution.

The size of the time step $\Dt$ does not change the error of explicit motor or
MFMD simulations (Figure~\ref{fig:5_performance}A), because  the steady-state
solution is time independent. The number of calculations increases linearly
with the number of time steps $N_t/\Dt$, making  the CPU time approximately
inversely proportional to $\Dt$. The MFMD error scales near-linearly with grid
spacing $\Ds$ as expected for a first-order upwind difference method
(Figure~\ref{fig:5_performance}B). The CPU time scales approximately as
$\Ds^{-2}$,  proportional to the number of grid points $N_{{\rm grid}} \propto
\Ds^{-2}$.

Explicit motor simulations have a cost that is linear in the motor number, but
the cost is constant for the MFMD and moment expansion models
(Figure~\ref{fig:5_performance}C). 
Fewer explicit motor simulations (24 realizations) were
needed to achieve sufficient statistics. We also note that at higher
concentration, the mean-field models return results closer to those of the
explicit model because stochastic fluctuations average out.  The explicit motor
model has a cost linear in filament length (due to the larger number of bound
motors on longer filaments), while for the MFMD model it is quadratic
(Figure~\ref{fig:5_performance}D).   The cost of the moment expansion model is
length independent. 

\section{Discussion}%
\label{sec:discussion}

To improve modeling methods for cytoskeletal filaments crosslinked by motors
(Figure~\ref{fig:1_overview}), we studied crosslinked filament pairs and
compared an explicit motor model to two levels of coarse-grained mean-field
motor models (Figure~\ref{fig:2_model_schematics}). The explicit motor model
uses Brownian dynamics and kinetic Monte Carlo to describe individual motor
binding and unbinding, motion, and force generation.  In the first level of
coarse graining, we average over individual motors and solve a PDE for the
mean-field motor density (MFMD). To further coarse grain, we compute a moment
expansion of the MFMD and solve a system of ODEs for the motor moments and
filament motion. 

We compared the model implementations for  filaments that are initially
antiparallel, parallel, or perpendicular
(Figure~\ref{fig:2-5_param_init_config}).  When filaments are held stationary,
the motor distribution reaches a steady state with similar average motor
distribution, force, and torque  for the three implementations
(Figure~\ref{fig:3_stat_compare}).  The explicit motor simulations showed
significant fluctuations that by construction are not present in the mean-field
models. Interestingly, we found that a significant portion of crosslinking
motors on antiparallel filaments do not reach their stall force for our
parameter set.

When filaments move, the final filament separation is similar for the explicit
motor and MFMD models, although the moment expansion model overestimates the
range of displacement and reorientation as a result of  neglecting boundary
terms (Figure~\ref{fig:4_cstr_compare}).  The dynamics of bound motor number,
force, and torque were similar for the MFMD and moment expansion models.  Motor
fluctuations in the explicit motor model lead to greater overall work done by
motors.

To compare computational cost across the model implementations, we studied
stationary filaments and motors at steady state
(Figure~\ref{fig:5_performance}). Both mean-field models have a simulation time
independent of motor concentration, potentially making them faster than
explicit models for systems with many motors. The moment expansion model's CPU
time is also independent of filament length, which could make it particularly
efficient for systems with long filaments. Overall, the moment expansion model
was $10^3-10^6$ faster than the other models. This method could therefore be
useful for simulating bulk active filament networks.

Future work could address the simplifying assumptions and approximations made
in the moment expansion model. An improved treatment of boundary terms may
improve the computation of filament motion. Incorporating additional motor
physics into the moment expansion model, such as non-zero length motors,
force-dependent detachment, and steric interactions between motors could
improve its ability to simulate microscopic motor behavior at the mesoscale,
bridging current explicit motor and continuous active network theories.
Implementing the moment expansion model in systems of many filaments is of
interest for testing whether the improvements in computational cost we identify
are present in larger systems.

\section*{Acknowledgements} This work was supported by NSF grants DMR-1725065
(MDB), DMS-1620003 (MAG and MDB), DMS-1620331(MJS), DMR-1420736 (MAG and ARL),
DMS-1463962(MJS),  and DMR-1420073 (MJS); NIH grant R01GM124371 (MDB); and a
fellowship provided by matching funds from the NIH/University of Colorado
Biophysics Training Program (ARL).  Simulations used the Summit supercomputer,
supported by NSF grants ACI-1532235 and ACI-1532236. 

\begin{appendices}

\section{Determining the time-step for binding}%
\label{sec:determining_proper_time_step_for_binding}
Our kinetic Monte Carlo algorithm assumes that multiple binding/unbinding
events do not occur in the same time step $\Dt$. As $\Dt$ becomes large
relative to the kinetic rates, this approximation fails. A time step is
appropriate if the maximum probability of two events occurring in  $\Dt$
satisfies
\begin{equation}
  \label{eq:max_prob}
 \max \{P(C(\Dt) \cup B(t')|A(0)) \} < \delta
\end{equation}
for a tolerance  $\delta$, where $A, B$, and $C$ denote motor bound states
(including unbound, single head bound, and crosslinking) at time $\Dt > t' >
0$. $P(C(\Dt) \cup B(t')|A(0))  = P(C(\Dt)|B(t'))P(B(t')|A(0))$ and each
individual state change follows a single event Poisson process with
$P(B(t)|A(0)) = 1 - \exp[-k_{A \to B} t]$.  The maximum probability for a
double event occurs at $t'=t'_{\rm{max}}$ found by solving
\begin{align}
\label{eq:t'_max}
\begin{split}
  \frac{dP(C(\Dt) \cup B(t')|A(0))}{dt'}\Bigg|_{t'_{max}} &= 0\\
  k_{A \to B} \left( e^{k_{B \to C}(\Dt-t')} - 1 \right)  -
  k_{B \to C} \left( e^{k_{A \to B} t'} - 1 \right)  &= 0.
\end{split}
\end{align}
While no analytic solution exists, $t'_{\rm{max}}$ can be numerically
computed.

There are four unique processes that must be considered with a two-step binding
process with unbound (U), single head bound (S), and crosslinking (C) states:
$U \to S \to U$,
$U \to S \to C$,
$S \to C \to S$, and 
$C \to S \to U$.
The process $C \to S \to C$ has the same probability as  $S \to C \to S$,
similarly,   $S \to U \to S$ has the same probability as $U \to S \to U$.  If
modeling filament motion with some force- or energy-dependent unbinding,
$\koff[,d]$ may be large. This means that in the limit of large unbinding rate
the probabilities $P(C \to S \to C) \to P(S \to C)$ and $P(C \to S \to U) \to
P(S \to U)$.

\section{Lookup table for kinetic Monte Carlo binding}
\label{sec:creating_lookup_table_for_kmc_rm_s_to_d_}
Equation (\ref{eq:KMC_SD_Eindep}) gives the transition probability of a singly
bound motor crosslinking as an integral of a Boltzmann factor. If $\hcl=0$,
$k_{{\rm on},C}$ is functionally similar to an error function. However, to
model non-zero-length tethers, we numerically integrate equation
(\ref{eq:KMC_SD_Eindep}).  Rather than directly numerically integrating at each
time step,  we precompute a lookup table.

The cumulative distribution function (CDF) of equation
(\ref{eq:KMC_SD_Eindep}), is a function $\hij$. All other variables in the
integral are constant for a given motor species.  We reduce the CDF
dimensionality by considering the lab position of each bound motor head and an
infinite carrier line defined by the position and orientation of the unbound
filament. Binding is then determined by the minimum distance $r_{\perp}$
between the bound motor head position and the filament ends $[s_-, s_+]$ on the
carrier line.

The carrier line CDF is
\begin{equation}
  \label{eq:CDF_inf}
  \CDF(r_{\perp},s) = \int_{-\infty}^{s} e^{-\beta U(r_{\perp}, s')}ds',
\end{equation}
allowing us to write the crosslinking rate as
\begin{equation}
  \label{eq:kon_LU}
  k_{{\rm on}, C}(r_{\perp}, s_+, s_-) = \ko[,d]\epsilon \Ke\left[ \CDF(r_{\perp}, s_+) - \CDF(r_{\perp}, s_-)\right].
\end{equation}
We notice that $e^{-\beta \Ucl}$ is symmetric in $s$, so $\CDF(r_{\perp},s) -
\CDF(r_{\perp},0)$ is anti-symmetric. Therefore, instead of integrating from
negative infinity, we use 
\begin{equation}
  \label{eq:CDF_as}
  \CDF'(r_{\perp},s) = \sgn(s)\int_{0}^{s} e^{-\beta U(r_{\perp}, s')}ds'
\end{equation}
and (\ref{eq:kon_LU}) to find the crosslinking rate.

We find the values of equation (\ref{eq:CDF_as}) by Gauss-Konrad integration.
The accuracy desired sets the maximum values for $s$ and $r_{\perp}$. The
integrand is always positive for real values of $s$ and $r_{\perp}$, so the CDF
asymptotes for large values of either variable. The maximum of the integral is
the point when the Boltzmann factor drops to the accuracy limit $\delta$.
Therefore, the lookup table domain is
\begin{equation}
  \label{eq:lut_limits}
  s, r_{\perp} \in \left[0, \sqrt{-\frac{2\ln(\delta)}{\beta \kcl}} + \hcl \right].
\end{equation}
Given a specified  grid spacing $\Delta s, \Delta r$, the  memory required for
the lookup table scales as $(s_{\max}/\Delta s) \times (r_{\perp, \max}/\Delta
r) $

\begin{figure}[htpb]
  \centering
  \includegraphics[width=1.0\linewidth]{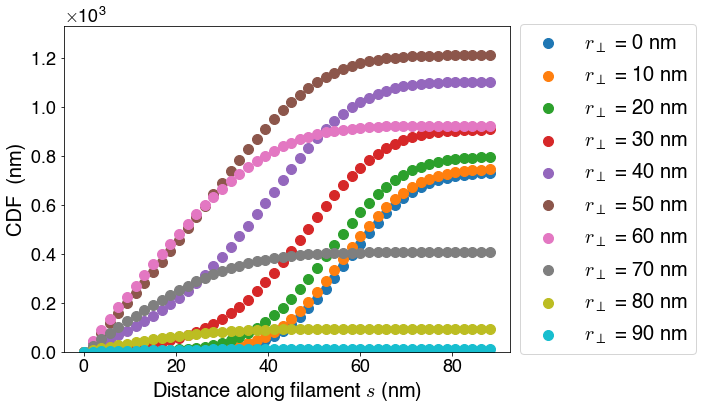}
  \caption{\footnotesize Visual representation of the lookup table showing CDF
    values as a function of distance $s$ along the filament for $\hcl = 32$ nm,
    $\kcl = .3$  pN/nm,  $\beta = 1./4.11$ (pN$\cdot$nm)$^{-1}$, and $\delta = 10^{-5}$.}
\label{fig:Lookup_table_example}
\end{figure}

\subsection{Interpolation of lookup table values}%
\label{sub:interpolation_of_values}
Since the lookup table is not a continuous function, we interpolate values
between discrete grid points. The 2D linear interpolation for input values of
$r_{\perp}$ and $s$ is
\begin{equation}
  \begin{split}
  \label{eq:2d_linear_interpolation}
  \CDF(r_{\perp}, s) \approx & \left( 1+m-\frac{r_{\perp}}{\Delta r} \right)\left( 1+n-\frac{s}{\Delta s} \right)\CDF_{m,n}
                    + \left( \frac{r_{\perp}}{\Delta r} - m \right)\left( 1+n-\frac{s}{\Delta s} \right)\CDF_{m+1,n}\\
                    +& \left( 1+m-\frac{r_{\perp}}{\Delta r} \right)\left( \frac{s}{\Delta s} - n \right)\CDF_{m,n+1}
                    + \left( \frac{r_{\perp}}{\Delta r} - m \right)\left( \frac{s}{\Delta s} - n \right)\CDF_{m+1,n+1},
  \end{split}
\end{equation}
where $\CDF_{m,n}=\CDF(m\Delta r, n\Delta s)$ are the lookup table
values at $m$ and $n$ if  $r_{\perp}$ lies within
$m\Delta r$ and $(m+1)\Delta r$ and $s$ lies within $n\Delta s$ and
$(n+1)\Delta s$.

\subsection{Reverse lookup algorithm}%
\label{sec:reverse_lookup_algorithm}
When a motor head binds to a filament, the binding position probability
distribution function (PDF) is defined by the Boltzmann factor.  We sample the
PDF by using the lookup table. To transform a uniform random variable $X$ to
random variable $Y$ with an arbitrary $\PDF_Y$, $X$ is inserted into the
inverted CDF of $Y$
\begin{equation}
  \label{eq:CDF_inversion}
  Y = \CDF_Y^{-1}(X).
\end{equation}
Since the lookup table holds the CDF values and given a random number from a
uniform distribution, we apply a combination of search and interpolation to
quickly find the corresponding random number from the PDF\@. The algorithm is
as follows
\begin{enumerate}\label{enum:lookup_alg}
  \item Sample a uniform random number $X \in [0, \CDF_{\max}]$. Note that the
    maximum value does not need to be 1.
  \item Given $r_{\perp}$, locate index $m$ such that $m\Delta r \leq r_{\perp}
    \leq (m +1) \Delta r$
  \item Use $m$ to find the set of indices $\{n_-, n_+\}$ such that
    $\CDF_{m,n_-} \leq X \leq \CDF_{m,n_-+1}$ and $\CDF_{m+1,n_+} \leq X \leq
    \CDF_{m+1,n_++1}$.
  \item Use the CDF values to interpolate the binding locations $s_-,
    s_+$ corresponding to the perpendicular distances $r_- = m\Delta r$ and
    $r_+ = (m+1) \Delta r$. For example,
    \begin{align}
      \label{eq:s_interp}
      s_- &= \Delta s \frac{X - \CDF_{m,n_-}}{\CDF_{m,n_-+1} - \CDF_{m,n_-}} + \Delta s n_-\\
      \label{eq:s_interp_+}
      s_+ &= \Delta s \frac{X - \CDF_{m+1,n_+}}{\CDF_{m+1,n_++1} - \CDF_{m+1,n_+}} + \Delta s n_+
    \end{align}
    Note that $s_-$ is not necessarily less than $s_+$.
  \item Find $s$ by interpolating the across the lookup table grid with
    respect to $r_{\perp}$
    \begin{equation}
      \label{eq:s_bound_avg}
      s \approx (s_+ - s_-)\frac{r_{\perp}-r_-}{\Delta r} + s_-
    \end{equation}
\end{enumerate}
While this algorithm succeeds in most circumstance, the low slope of the CDF at
large values of $s$ can cause errors. For example, if the lookup table has the
form of Figure~\ref{fig:Lookup_table_example} and a protein is located at a
perpendicular distance of $r_{\perp}=35$ nm, given a random number of $X=10^3$,
no value for $s_-$ will be found since $\CDF(30, s_{\rm max}) < 10^3$.  To
correct for this, we solve for $s$ using a binary search algorithm. 

The binary search algorithm is as follows
\begin{enumerate}\label{enum:bin_search_alg}
    \item Determine if $\CDF_{m,n_{\rm max}}$ or $\CDF_{m+1,n_{\rm max}}$ is
      less than $X$. If $\CDF_{m,n_{\rm max}} < X$, set $s_- =s_{\rm max}$. If
      $\CDF_{m+1,n_{\rm max}} < X$, set $s_+ =s_{\rm max}$.
    \item Find other $s_\pm$ using the inverted lookup table and equation
      (\ref{eq:s_interp}) or (\ref{eq:s_interp_+}).
    \item Find the average of $s_-$ and $s_+$.
    \item Use the lookup table interpolation algorithm to find the
      $\CDF(r_{\perp}, s_{\rm avg})$.
    \item If $\CDF(r_{\perp}, s_{\rm avg}) > X$ set the larger of the two
      $s_{\pm}$ values to $s_{\rm avg}$. Otherwise, set the smaller of the two
      to $s_{\rm avg}$.
    \item Repeat steps 3-5 until $|\CDF(r_{\perp}, s_{\rm avg}) - X| < \delta$
      for some desired tolerance $\delta$.
\end{enumerate}
This process converges at a rate $O(\log_2(\delta s_{\max}))$. 



\section{Numerical integration of the MFMD equation}%
\label{sec:numerical_solutions}
We approximate the solution $\psij(s_i, s_j, t)$ by discretizing the solution
in time and space
\begin{equation}
  \label{eq:discrete_psi}
  \psij(s_i, s_j, t) \to \PSI = \psij(m\Ds , n\Ds , k\Dt)
\end{equation}
for $\PSI{} \in \mathbb{R}^{(M_i+1) \times (M_j+1) \times k}$, where $M_i$ is
the number of discretized points along filament $i$. Additional boundary points
for $m,n=0$ are added.

We use forward Euler time-stepping so our discrete differential operator for
time is
\begin{equation}
  \label{eq:f_euler_time}
  \pder[\psij]{t} \to \frac{1}{\Dt}(\PSI - \PSI[m][n][k-1]).
\end{equation}

To solve the hyperbolic FPE (\ref{eq:fpe}), we use a first-order accurate
upwind method\cite{leveque07a}. The differential operator for $s_i$ becomes
\begin{equation}
  \label{eq:upwind_op}
  \pder[\psij]{s_i} \to \frac{1}{\Ds}(\PSI - \PSI[m-1]).
\end{equation}
Note this only holds for the indices $0<m$ and $0<n$. The matrix
representation for equation (\ref{eq:upwind_op}) is
\begin{equation}
  \label{eq:matrix_D}
  \frac{1}{\Delta s}
  \begin{pmatrix}
    c_0 & c_1 & c_2 & \cdots & c_{M_i}\\
    -1 & 1 & 0 & &\\
    0 & -1 & 1& \ddots &\\
    \vdots & & \ddots & \ddots & 0\\
    d_0& \cdots & & d_{M_i-1} & d_{M_i}
  \end{pmatrix}
  \begin{pmatrix}
    \PSI[0]{}\\\PSI[1]{}\\\vdots\\\PSI[M_i-1]{}\\\PSI[M_i]{}
  \end{pmatrix}
  = \rhd^{m,a} \PSI[a],
\end{equation}
where $c_m$ and $d_m$ are chosen to satisfy the boundary conditions. We choose
the notation $\rhd^{m,n}$ for this matrix. To differentiate along $s_j$, we use
the identity $\PSI = \psi_{j,i}^{n,m,k}$, apply $\rhd^{m,n}$ on the matrix, and
then convert back,
\begin{equation}
  \label{eq:matrix_D_j}
  \pder[\psij]{s_i} \to \left(\sum_a \rhd^{n,a} \psi_{j,i}^{a,m} \right)^T,
\end{equation}
which in index notation is $\psij^{m,a}(\rhd^T)^{n,a}$. For
brevity, we use the notation $\psij^{m,a}(\rhd^T)^{n,a}=\psij^{m,a}\lhd^{a,n}$.

The discretized Fokker-Planck equation (\ref{eq:fpe}) is then
\begin{equation}
  \label{eq:discrete_motor_fpe}
  \PSI[m][n][k+1] = \Dt \left(-\rhd^{m,a} \left(v_{i,j}^{a,n,k}\PSI[a]\right)
    -\left(v_{j,i}^{m,a,k}\PSI[m][a] \right)\lhd^{a,n} + 2\ko c e^{-\beta U_{i,j}^{m,n,k}} -
    2\ko\PSI\right) + \PSI{},
\end{equation}
where $U_{i,j}^{m,n,k}$ and $v_{i,j}^{m,n,k}$ are the discretized potential and
velocity at time $k\Dt$. Note that $U_{i,j}^{m,n,k} = U_{j,i}^{n,m,k}$, but
$v_{i,j}^{m,n,k} \neq v_{j,i}^{n,m,k}$.

In cases where the flux of the motors $\pder[\left(v_{i,j}\psij\right)]{s_i}$
is known at the boundaries, we construct $\rhd$ to satisfy the requirements.
When filaments are in solution, there is zero flux from the minus ends, so all
$c_m = 0$. In our simulations, motors walk of filament ends with out pausing,
so $d_{M_i-1}=-1$ and $d_{M_i}=1$ with all other $d_m=0$. Although not modeled
in this paper, some biological motors end pause at filament plus ends. To model
this, $d_{M_i-1}=-1$ and every other $d_m=0$.

\section{Conversion of binding parameters from an explicit to mean-field motor density model}%
\label{sec:conversion_of_binding_parameters}

To relate binding parameters of the one-step and multi-step binding models, we
use that at steady state, the motor distribution $\psij$ should be equivalent
for both models.  Since we only compare binding kinetics, we simplify the
Fokker-Planck equation to keep only the binding terms: in equation
(\ref{eq:fpe}), we set $v_{i,j} = v_{j,i} = 0$, 
\begin{equation}
  \label{eq:static_fpe}
  \pder[\psij]{t} = 2\ko c e^{-\beta \Ucl} - 2\ko \psij.
\end{equation}
The steady-state solution is
\begin{equation}
  \label{eq:static_fpe_ss}
  \psij= c e^{-\beta \Ucl},
\end{equation}
which is a Boltzmann factor multiplied by an effective concentration.

The multi-step binding model can be written
\begin{align}
  \begin{split}
  \label{eq:multi_psij_static}
  \frac{\del \psij(s_i, s_j)}{\del t} &= \epsilon \Ke' \ko[,C](\chi_i+\chi_j)e^{-\beta \Ucl} - 2\ko[,C] \psij,
  \end{split}\\
  \begin{split}
    \label{eq:chi_i_static}
  \frac{\del \chi_i(s_i)}{\del t} &= c_o K_a \epsilon \ko[,S] - \ko[,S]\chi_i
  + \int_{L_j} \left( \ko[,C]\psij - \epsilon \Ke' \ko[,C]\chi_i e^{-\beta \Ucl} \right) ds_j ,
  \end{split}\\
  \begin{split}
    \label{eq:chi_j_static}
  \frac{\del \chi_j(s_j)}{\del t} &= c_o K_a \epsilon \ko[,S] - \ko[,S]\chi_j
  + \int_{L_i} \left( \ko[,C]\psij - \epsilon \Ke' \ko[,C]\chi_j e^{-\beta \Ucl} \right) ds_i ,
  \end{split}
\end{align}
where $\chi_i$ is the mean-field density of motors with one head bound to
filament $i$ (cf.  Equation~\ref{eq:explicit_chi}).  We define $\Ke' =
\Ke/\Vbd$ and solve for the steady state, giving
\begin{align}
  \begin{split}
  \label{eq:mulit_psij_static_ss}
    \psij &= \frac{\epsilon \Ke'}{2}(\chi_i+\chi_j)e^{-\beta \Ucl},
  \end{split}\\
  \begin{split}
    \label{eq:chi_i_static_ss}
    \chi_i &= \frac{\epsilon \Ke' \ko[,C]}{2 \ko[,S]} \left(\int_{L_j} (\chi_j-\chi_i) e^{-\beta \Ucl}ds_j\right) + \epsilon K_a c_o,
  \end{split}\\
  \begin{split}
    \label{eq:chi_j_static_ss}
    \chi_j &= \frac{\epsilon \Ke' \ko[,C]}{2 \ko[,S]} \left(\int_{L_i} (\chi_i-\chi_j) e^{-\beta \Ucl}ds_i\right) + \epsilon K_a c_o.
  \end{split}
\end{align}
The equations for $\chi_i$ and $\chi_j$ have the 
 forms
\begin{align}
  \begin{split}
    \label{eq:X}
    X(s) &= C\int_{a}^{b} \left(Y(t)-X(s)\right) K(s,t)dt + D,
  \end{split}\\
  \begin{split}
    \label{eq:Y}
    Y(t) &= C\int_{c}^{d} \left(X(s)-Y(t)\right) K(s,t)ds + D,
  \end{split}
\end{align}
where $t \in [a,b]$ and $s \in [c,d]$. Distributing the integrals, we can
rewrite
\begin{align}
  \begin{split}
    \label{eq:X_dist}
    X(s) &= C\int_{a}^{b} Y(t)K(s,t)dt - CX(s)F(s)  + D,
  \end{split}\\
  \begin{split}
    \label{eq:Y_dist}
    Y(t) &= C\int_{c}^{d} X(s) K(s,t)ds - CY(t)G(t) + D,
  \end{split}
\end{align}
where $F(s) = \int_{a}^{b}K(s,t)dt$ and $G(t) = \int_{c}^{d}K(s,t)ds$. Solving
for $X(s)$ and $Y(t)$ gives
\begin{align}
  \begin{split}
    \label{eq:X_s}
    X(s) &= \frac{D}{1+CF(s)} + \frac{C}{1+CF(s)}\int_{a}^{b} Y(t) K(s,t)dt,
  \end{split}\\
  \begin{split}
    \label{eq:Y_t}
    Y(t) &= \frac{D}{1+CG(t)} + \frac{C}{1+CG(t)}\int_{c}^{d} X(s) K(s,t)ds,
  \end{split}
\end{align}
After plugging equation (\ref{eq:X_s}) into (\ref{eq:Y_t}) we find
\begin{equation}
  \label{eq:Y_X}
  Y(t) = \frac{D}{1+CG(t)} + \frac{CD}{1+CG(t)}\int_{c}^{d}\frac{K(s,t)}{1+CF(s)}ds + C^2\int_{a}^{b}\int_{c}^{d}Y(t') \frac{K(s,t)K(s,t')}{(1+CF(s))(1+CG(t'))} ds dt'.
\end{equation}
This can be rearranged into the form
\begin{equation}
  \label{eq:Y_X_simp}
  Y(t) = A(t) + \int_{a}^{b}Y(t') B(t,t') dt',
\end{equation}
which implies that $Y(t)$ and $X(s)$ each satisfy a Fredholm equation of the
second kind.  Both $A$ and $B$ are continuous given $K(s,t)= e^{-\beta
  \Ucl(s,t)}$, so the Fredholm equations of the second kind have unique
solutions. By inspection, the solution to equations (\ref{eq:X}) and
(\ref{eq:Y}) is $X(s)=Y(t)=D$. When we substitute this solution in equations
(\ref{eq:chi_i_static_ss}) and (\ref{eq:chi_j_static_ss}), we find $\chi_i =
\chi_j = \epsilon K_a c_o$ and
\begin{equation}
  \label{eq:static_fpe_ss_approx}
    \psij = \epsilon^2 K_a \Ke'c_o e^{-\beta \Ucl}.
\end{equation}
Setting equation (\ref{eq:static_fpe_ss}) equal to
(\ref{eq:static_fpe_ss_approx}) gives
\begin{equation} 
  \label{eq:concentration_compare_2} 
  c = \frac{\epsilon^2 K_a \Ke}{\Vbd}c_o.
\end{equation}

\section{Calculating binding parameters from experiments}%
\label{sub:calculating_binding_parameters_from_experiments}
The experimental parameters for motor binding are not always independently
measured. If all but one binding parameters are known, then the unknown
parameter can be found from equation (\ref{eq:concentration_compare}) and the
ratio of the number of motors with one head bound and number of motors
crosslinking. 

As an example, suppose we wish to find  $\Ke$.  The number of motors with one
head bound is $N_S = c_o K_a \epsilon L$, where $L$ is the filament length.
\textit{In vitro} experiments~\cite{shimamoto15} can measure the crosslinking
motors number $N_d$. Integrating equation (\ref{eq:static_fpe_ss_approx}), we
the model prediction for the number of crosslinking motors is \begin{equation}
  \label{eq:n_d} N_C =  c_o \epsilon^2 K_a \Ke' \int_{L_{i}} \int_{L_{j}}
  e^{-\beta \Ucl} ds_i ds_j.  \end{equation} For fully  parallel or
antiparallel filaments of the same length with adjacent centers, the total
number of motors in equation (\ref{eq:n_d}) is proportional to $L$.  If $L \gg
\sqrt{2/\beta \kcl}$, the Gaussian integral $\approx L\sqrt{2\pi/\beta
  \kcl}e^{-\beta \kcl r_{\perp}^2}$, where $r_{\perp}$ is the center-to-center
separation between filaments. The ratio of the number of crosslinking motors
relative to the number motors with one head bound is
\begin{equation}
  \label{eq:ratio_s_to_d}
  \rho = \frac{N_C}{N_S} = \epsilon \Ke' \sqrt{\frac{2\pi}{\beta \kcl}}e^{-\beta \kcl r_{\perp}^2},
\end{equation}
allowing us to estimate 
$\Ke' = \frac{\rho}{\epsilon}\sqrt{\frac{\beta \kcl}{2\pi}}e^{\beta \kcl
  r_{\perp}^2}$.

\section{Gaussian integrals in the moment expansion}%
\label{sec:speed_up_of_gaussian_integrals_in_moment_expansion}
The source terms in the moment expansion require a double integral over two
filaments. To lower the numerical integration's computational cost, we find an
analytic solution for either the semi-integrated term $Q_j^l(s_i)$ or the
fully-integrated term $q_{i,j}^{k,l}$. 

The integrated source terms are
\begin{align}
  \label{eq:qk,l}
  \begin{split}
  q_{i,j}^{k,l} = c e^{-\left(\frac{r}{\alpha}\right)^2}
  &\Bigg\{ \int_{L_i} s_i^k \exp\left[-\frac{s_i^2 - 2s_i\ru - (\ru[j][i] + \cb s_i)^2}{\alpha^2}\right] \\
    &\int_{L_j} s_j^l \exp\left[-\left(\frac{s_j - \ru[j][i] - \cb s_i}{\alpha}\right)^2\right] ds_j ds_i\Bigg\},
   \end{split}
 \end{align}
where $\alpha = \sqrt{\frac{2}{\beta\kcl}}$. We define the
quantity $A=-\ru[j][i]- \cb s_i$ so that the integral over $s_j$ becomes
\begin{equation}
  \label{eq:Ql_simp}
  \bar{Q}^l_j(s_i) = \int_{L_j} s_j^l e^{-{\left(\frac{s_j+A}{\alpha}\right)}^2}ds_j.
\end{equation}
This integral has an analytic form in terms of
error functions, which can be rapidly computed. For $l={0,1,2,3}$, we find
\begin{equation}
  \label{eq:Q0}
  \bar{Q}_j^0(s_i) =\frac{\alpha \sqrt{\pi}}{2}
              \left[ \erf \left( \frac{s_j+A}{\alpha} \right) \right]_{\partial L_j}
\end{equation}
\begin{equation}
  \label{eq:Q1}
  \bar{Q}_j^1(s_i) = -\frac{\alpha }{2}\left[\alpha e^{-\left(\frac{s_j+A}{\alpha}\right)^2}
              + A\sqrt{\pi}\erf \left( \frac{s_j+A}{\alpha} \right) \right]_{\partial L_j}
\end{equation}
\begin{equation}
  \label{eq:Q2}
  \bar{Q}_j^2(s_i) = \frac{\alpha}{4}\left[ 2\alpha(A-s_j) e^{-\left(\frac{s_j+A}{\alpha}\right)^2}
    + (2A^2+\alpha^2)\sqrt{\pi}\erf \left( \frac{s_j+A}{\alpha}\right) \right]_{\partial L_j}
\end{equation}
\begin{equation}
  \label{eq:Q3}
  \bar{Q}_j^3(s_i) = \frac{-\alpha}{4}\left[ 2\alpha(A^2 - A s_j + s_j^2 + \alpha^2)
                        e^{-\left(\frac{s_j+A}{\alpha}\right)^2}
              + (2A^2 + 3\alpha^2)A\sqrt{\pi}\erf \left( \frac{s_j+A}{\alpha}\right)
            \right]_{\partial L_j}
\end{equation}

\section{Moment expansion boundary terms}%
\label{sub:boundary_value_calculations}

To generally define boundary conditions, instead of integrating over both $s_i$
and $s_i$, we integrate over just one variable. This makes  the boundary
condition a function of a single filament attatchment position. For example,
the boundary terms for the first filament are
\begin{align}
  \label{eq:boundary_term_gen}
  \begin{split}
  \dot{B}_j^l(s_i) = \int_{L_j} & \left( 2\ko c e^{-\beta U_{i,j}}
      +(2\kappa - 2\ko)\psij
      - \left( v_o +\kappa(\ru + \cb s_j - s_i) \right)\pder[\psij]{s_i}\right.\\
      &\left.-\left( v_o+\kappa(\ru[j][i] + \cb s_i - s_j ) \right)\pder[\psij]{s_j} \right)
      s_j^l ds_j.
  \end{split}
\end{align}
These boundary terms are evaluated at $-L_i/2$ and $L_i/2$. We derive a
recursion relation by integrating equation (\ref{eq:boundary_term_gen}) over
$s_j$ and using the definition in equation (\ref{eq:bjl}) 
\begin{align}
  \label{eq:boundary_evol_j}
  \begin{split}
    \dot{B}_j^l\left(s_i\right) &=  2\ko cQ_j^l\left(s_i\right)
    + l \left( v_o + \kappa (\ru[j][i] + \cb s_i)\right)B_j^{l-1}
    - \left(2\ko + \kappa(l-1)\right)B_j^l\\
    &- \left( v_o + \kappa(\ru[i,j] - s_i) \right)\pder[B_{j}^{l}]{s_i}
    - \kappa \cb \pder[B_{j}^{l+1}]{s_i} \\
    &- \left[ s_j^l \left(v_o + \kappa \left( \ru[j][i] + \cb s_i - s_j\right) \right)
      \psij\left(s_i, s_j\right) \right]_{\del L_j}.
  \end{split}
\end{align}
Solving this equation requires finding the time evolution of the boundary term
spatial derivatives, which solve
\begin{align}
  \label{eq:der_boundary_evol_j}
  \begin{split}
    \pder[\dot{B}_j^l\left(s_i\right)]{s_i} &=  2\ko c\pder[Q_j^l]{s_i}
     + \cb l B_j^{l-1}\\
    &+ l \left( v_o + \kappa (\ru[j][i] + \cb s_i)\right)\pder[B_j^{l-1}]{s_i}
     - \left(2\ko + \kappa(l-2)\right)\pder[B_j^l]{s_i}\\
    &- \left( v_o + \kappa(\ru[i,j] - s_i) \right)\pdertwo[B_{j}^{l}]{s_i}
    + \kappa \cb \pdertwo[B_{j}^{l+1}]{s_i} \\
    &+ \textrm{corner terms}.
  \end{split}
\end{align}
This shows that the boundary terms do not close. However, if the higher-order
terms or their coefficients are small compared to the moments
$\mu_{i,j}^{k,l}$, we may take a zeroth-order approximation. We consider this
approximation in Section~\ref{sec:results}.

\end{appendices}
\bibliographystyle{unsrt}
\bibliography{bibliography_2, zoterolibrary, active_stress}

\end{document}

%% file: params_units_tab.tex
\begin{table}[t!]
\centering
  \footnotesize
  \begin{tabular}{lllp{4.5cm}}
  \textbf{Parameter} & \textbf{Symbol} & \textbf{Value} & \textbf{Notes}\\
  Total time & $N_t$ & 20 $\rm sec$  & Chosen \\
  Explicit motor time step size & $\Delta t_{\rm{Explicit}}$ & 0.0001 $\rm sec$  & Chosen for numerical stability \\
  MFMD time step size & $\Delta t_{\rm{MFMD}}$ & 0.001 $\rm sec$  & Chosen for numerical stability \\
  MFMD grid spacing & $\Delta s$ & 1 $\rm nm$  & Chosen for numerical stability \\
  Viscosity & $\eta$ & $10^{-6}$ $\rm pN$ $\rm sec$ $\rm nm^{-2}$  & Chosen (viscosity of cytoplasm) \\
  Filament length & L & 1 $\rm \mu m$  & Chosen \\
  Filament diameter & D$_{\rm{fil}}$ & 25 $\rm nm$  & Diameter of microtubules \cite{alberts08} \\
  Explicit motor concentration & $c_o$ & 11 $\rm nM$  & Chosen \\
  MFMD effective concentration & $c$ & 0.0093 $\rm nm^{-2}$  & Calculated (Section \ref{sec:conversion_of_binding_parameters}) \\
  Modified tether length & $h_{\rm{cl}}$ & 0 $\rm nm$  & Chosen (Section \ref{sub:motors}) \\
  Effective tether \\ spring constant & $k_{\rm{cl}}$ & 0.037 $\rm pN$ $\rm nm^{-1}$  & Calculated (Section \ref{sub:motors}), spring constant \cite{kawaguchi01}, tether length \cite{kashina96} \\
  Filament binding site density & $\epsilon$ & 0.25 $\rm nm^{-1}$  & Estimated, one site every four nanometers \\
  Inverse temperature & $\beta=\frac{1}{k_B T}$ & 0.2433 $\rm pN^{-1}$ $\rm nm^{-1}$  & Room temperature \\
  Motor speed & $v_o$ & 50 $\rm nm$ $\rm sec^{-1}$  & \cite{gerson-gurwitz11} \\
  Motor stall force & $f_{stall}$ & 2 $\rm pN$  & \cite{valentine06} \\
  Association constant \\ (unbound$\leftrightarrow$one head bound) & $K_{a}$ & 0.005 $\rm nM^{-1}$  & \cite{cochran15} \\
  Association constant \\ (one head bound$\leftrightarrow$crosslinking) & $K'_{e}$ & 2.56  & Calculated (Section \ref{sec:conversion_of_binding_parameters}), \cite{shimamoto15} \\
  Multi-step bare off rate \\ (unbound$\leftrightarrow$one head bound) & $\ko[,S]$ & 0.77 $\rm sec^{-1}$  & \cite{cochran15} \\
  Multi-step bare off rate \\ (one head bound$\leftrightarrow$crosslinking) & $\ko[,C] $ & 0.77 $\rm sec^{-1}$  & Chosen to match $\ko[,S]$ \\
  One-step bare off rate & $\ko$ & 0.77 $\rm sec^{-1}$  & Chosen to match $\ko[,S]$ \\
 \end{tabular}
 \caption{\footnotesize Model parameters for MTs and kinesin-5 for explicit motor distribution and MFMD calculations.}\label{tab:units}
\end{table}

%% file: motor_methods_main.bbl
\begin{thebibliography}{10}

\bibitem{howard01}
Jonathon Howard.
\newblock {\em Mechanics of Motor Proteins and the Cytoskeleton}.
\newblock {Sinauer Associates, Publishers}, {Sunderland, Mass}, 2001.

\bibitem{bray01}
Dennis Bray.
\newblock {\em Cell Movements: From Molecules to Motility}.
\newblock {Garland Pub}, {New York}, 2nd ed edition, 2001.

\bibitem{blanchoin14}
Laurent Blanchoin, Rajaa {Boujemaa-Paterski}, C{\'e}cile Sykes, and Julie
  Plastino.
\newblock Actin {{Dynamics}}, {{Architecture}}, and {{Mechanics}} in {{Cell
  Motility}}.
\newblock {\em Physiological Reviews}, 94(1):235--263, January 2014.

\bibitem{pollard09}
Thomas~D. Pollard and John~A. Cooper.
\newblock Actin, a {{Central Player}} in {{Cell Shape}} and {{Movement}}.
\newblock {\em Science}, 326(5957):1208--1212, November 2009.

\bibitem{mcintosh12}
J.~Richard McIntosh, Maxim~I. Molodtsov, and Fazly~I. Ataullakhanov.
\newblock Biophysics of mitosis.
\newblock {\em Quarterly Reviews of Biophysics}, 45(2):147--207, May 2012.

\bibitem{huxley57}
AF. HUXLEY.
\newblock Muscle structure and theories of contraction.
\newblock {\em Prog. Biophys. Biophys. Chem}, 7:255--318, 1957.

\bibitem{huxley96}
A.~F. Huxley and S.~Tideswell.
\newblock Filament compliance and tension transients in muscle.
\newblock {\em Journal of Muscle Research \& Cell Motility}, 17(4):507--511,
  August 1996.

\bibitem{huxley97}
A.~F. HUXLEY and S.~TIDESWELL.
\newblock Rapid regeneration of power stroke in contracting muscle by
  attachment of second myosin head.
\newblock {\em Journal of Muscle Research \& Cell Motility}, 18(1):111--114,
  February 1997.

\bibitem{gupton06}
Stephanie~L. Gupton and Clare~M. {Waterman-Storer}.
\newblock Spatiotemporal {{Feedback}} between {{Actomyosin}} and
  {{Focal}}-{{Adhesion Systems Optimizes Rapid Cell Migration}}.
\newblock {\em Cell}, 125(7):1361--1374, June 2006.

\bibitem{fournier10}
Maxime~F. Fournier, Roger Sauser, Davide Ambrosi, Jean-Jacques Meister, and
  Alexander~B. Verkhovsky.
\newblock Force transmission in migrating cells.
\newblock {\em Journal of Cell Biology}, 188(2):287--297, January 2010.

\bibitem{barnhart11}
Erin~L. Barnhart, Kun-Chun Lee, Kinneret Keren, Alex Mogilner, and Julie~A.
  Theriot.
\newblock An {{Adhesion}}-{{Dependent Switch}} between {{Mechanisms That
  Determine Motile Cell Shape}}.
\newblock {\em PLOS Biology}, 9(5):e1001059, May 2011.

\bibitem{laevsky03}
G.~Laevsky.
\newblock Cross-linking of actin filaments by myosin {{II}} is a major
  contributor to cortical integrity and cell motility in restrictive
  environments.
\newblock {\em Journal of Cell Science}, 116(18):3761--3770, September 2003.

\bibitem{hagan92}
Iain Hagan and Mitsuhiro Yanagida.
\newblock Kinesin-related cut 7 protein associates with mitotic and meiotic
  spindles in fission yeast.
\newblock {\em Nature}, 356(6364):74, March 1992.

\bibitem{saunders95}
W.~S. Saunders, D.~Koshland, D.~Eshel, I.~R. Gibbons, and M.~A. Hoyt.
\newblock Saccharomyces cerevisiae kinesin- and dynein-related proteins
  required for anaphase chromosome segregation.
\newblock {\em Journal of Cell Biology}, 128(4):617--624, February 1995.

\bibitem{kapoor00}
Tarun~M. Kapoor, Thomas~U. Mayer, Margaret~L. Coughlin, and Timothy~J.
  Mitchison.
\newblock Probing {{Spindle Assembly Mechanisms}} with {{Monastrol}}, a {{Small
  Molecule Inhibitor}} of the {{Mitotic Kinesin}}, {{Eg5}}.
\newblock {\em Journal of Cell Biology}, 150(5):975--988, September 2000.

\bibitem{cai08}
Shang Cai, Lesley~N. Weaver, Stephanie~C. {Ems-McClung}, and Claire~E. Walczak.
\newblock Kinesin-14 {{Family Proteins HSET}}/{{XCTK2 Control Spindle Length}}
  by {{Cross}}-{{Linking}} and {{Sliding Microtubules}}.
\newblock {\em Molecular Biology of the Cell}, 20(5):1348--1359, December 2008.

\bibitem{wollman08}
Roy Wollman, Gul {Civelekoglu-Scholey}, Jonathan~M Scholey, and Alex Mogilner.
\newblock Reverse engineering of force integration during mitosis in the
  {{Drosophila}} embryo.
\newblock {\em Molecular Systems Biology}, 4(1):195, January 2008.

\bibitem{civelekoglu-scholey10a}
Gul {Civelekoglu-Scholey} and Jonathan~M. Scholey.
\newblock Mitotic force generators and chromosome segregation.
\newblock {\em Cellular and Molecular Life Sciences}, 67(13):2231--2250, July
  2010.

\bibitem{she17}
Zhen-Yu She and Wan-Xi Yang.
\newblock Molecular mechanisms of kinesin-14 motors in spindle assembly and
  chromosome segregation.
\newblock {\em Journal of Cell Science}, 130(13):2097--2110, July 2017.

\bibitem{vukusic17}
Kruno Vuku{\v s}i{\'c}, Renata Buda, Agneza Bosilj, Ana Milas, Nenad Pavin, and
  Iva~M. Toli{\'c}.
\newblock Microtubule {{Sliding}} within the {{Bridging Fiber Pushes
  Kinetochore Fibers Apart}} to {{Segregate Chromosomes}}.
\newblock {\em Developmental Cell}, 43(1):11--23.e6, October 2017.

\bibitem{ganguly12}
Sujoy Ganguly, Lucy~S. Williams, Isabel~M. Palacios, and Raymond~E. Goldstein.
\newblock Cytoplasmic streaming in {{Drosophila}} oocytes varies with kinesin
  activity and correlates with the microtubule cytoskeleton architecture.
\newblock {\em Proceedings of the National Academy of Sciences},
  109(38):15109--15114, September 2012.

\bibitem{satir68}
Peter Satir.
\newblock {{STUDIES ON CILIA}}.
\newblock {\em The Journal of Cell Biology}, 39(1):77--94, October 1968.

\bibitem{summers71}
Keith~E. Summers and I.~R. Gibbons.
\newblock Adenosine {{Triphosphate}}-{{Induced Sliding}} of {{Tubules}} in
  {{Trypsin}}-{{Treated Flagella}} of {{Sea}}-{{Urchin Sperm}}.
\newblock {\em Proceedings of the National Academy of Sciences of the United
  States of America}, 68(12):3092--3096, December 1971.

\bibitem{king18}
Stephen~M. King.
\newblock Turning dyneins off bends cilia.
\newblock {\em Cytoskeleton (Hoboken, N.j.)}, 75(8):372--381, August 2018.

\bibitem{nedelec97}
F.~J. Nedelec, T.~Surrey, A.~C. Maggs, and S.~Leibler.
\newblock Self-organization of microtubules and motors.
\newblock {\em Nature}, 389(6648):305--308, September 1997.

\bibitem{surrey01}
Thomas Surrey, Fran{\c c}ois N{\'e}d{\'e}lec, Stanislas Leibler, and Eric
  Karsenti.
\newblock Physical {{Properties Determining Self}}-{{Organization}} of
  {{Motors}} and {{Microtubules}}.
\newblock {\em Science}, 292(5519):1167--1171, May 2001.

\bibitem{Backouche2006}
F.~Backouche, L.~Haviv, D.~Groswasser, and A.~Bernheim-Groswasser.
\newblock {Active gels: dynamics of patterning and self-organization}.
\newblock {\em Physical Biology}, 3(4):264--273, dec 2006.

\bibitem{sanchez12a}
Tim Sanchez, Daniel T.~N. Chen, Stephen~J. DeCamp, Michael Heymann, and
  Zvonimir Dogic.
\newblock Spontaneous motion in hierarchically assembled active matter.
\newblock {\em Nature}, 491(7424):431--434, November 2012.

\bibitem{doostmohammadi18}
Amin Doostmohammadi, Jordi {Ign{\'e}s-Mullol}, Julia~M. Yeomans, and Francesc
  Sagu{\'e}s.
\newblock Active nematics.
\newblock {\em Nature Communications}, 9(1):3246, August 2018.

\bibitem{lemma19}
Linnea~M. Lemma, Stephen~J. DeCamp, Zhihong You, Luca Giomi, and Zvonimir
  Dogic.
\newblock Statistical properties of autonomous flows in {{2D}} active nematics.
\newblock {\em Soft Matter}, 15(15):3264--3272, April 2019.

\bibitem{duclos20}
Guillaume Duclos, Raymond Adkins, Debarghya Banerjee, Matthew S.~E. Peterson,
  Minu Varghese, Itamar Kolvin, Arvind Baskaran, Robert~A. Pelcovits, Thomas~R.
  Powers, Aparna Baskaran, Federico Toschi, Michael~F. Hagan, Sebastian~J.
  Streichan, Vincenzo Vitelli, Daniel~A. Beller, and Zvonimir Dogic.
\newblock Topological structure and dynamics of three-dimensional active
  nematics.
\newblock {\em Science}, 367(6482):1120--1124, March 2020.

\bibitem{brugues12}
Jan Brugu{\'e}s, Valeria Nuzzo, Eric Mazur, and Daniel~J. Needleman.
\newblock Nucleation and {{Transport Organize Microtubules}} in {{Metaphase
  Spindles}}.
\newblock {\em Cell}, 149(3):554--564, April 2012.

\bibitem{roostalu18}
Johanna Roostalu, Jamie Rickman, Claire Thomas, Fran{\c c}ois N{\'e}d{\'e}lec,
  and Thomas Surrey.
\newblock Determinants of {{Polar}} versus {{Nematic Organization}} in
  {{Networks}} of {{Dynamic Microtubules}} and {{Mitotic Motors}}.
\newblock {\em Cell}, 175(3):796--808.e14, October 2018.

\bibitem{weirich19}
Kimberly~L. Weirich, Kinjal Dasbiswas, Thomas~A. Witten, Suriyanarayanan
  Vaikuntanathan, and Margaret~L. Gardel.
\newblock Self-organizing motors divide active liquid droplets.
\newblock {\em Proceedings of the National Academy of Sciences},
  116(23):11125--11130, June 2019.

\bibitem{nedelec07}
Francois Nedelec and Dietrich Foethke.
\newblock Collective {{Langevin}} dynamics of flexible cytoskeletal fibers.
\newblock {\em New Journal of Physics}, 9(11):427, November 2007.

\bibitem{popov16}
Konstantin Popov, James Komianos, and Garegin~A. Papoian.
\newblock {{MEDYAN}}: {{Mechanochemical Simulations}} of {{Contraction}} and
  {{Polarity Alignment}} in {{Actomyosin Networks}}.
\newblock {\em PLOS Computational Biology}, 12(4):e1004877, April 2016.

\bibitem{freedman17}
Simon~L. Freedman, Shiladitya Banerjee, Glen~M. Hocky, and Aaron~R. Dinner.
\newblock A {{Versatile Framework}} for {{Simulating}} the {{Dynamic Mechanical
  Structure}} of {{Cytoskeletal Networks}}.
\newblock {\em Biophysical Journal}, 113(2):448--460, July 2017.

\bibitem{head14}
D.~A. Head, W.~J. Briels, and Gerhard Gompper.
\newblock Nonequilibrium structure and dynamics in a microscopic model of
  thin-film active gels.
\newblock {\em Physical Review E}, 89(3):032705, March 2014.

\bibitem{aranson05}
Igor~S. Aranson and Lev~S. Tsimring.
\newblock Pattern formation of microtubules and motors: {{Inelastic}}
  interaction of polar rods.
\newblock {\em Physical Review E}, 71(5), May 2005.

\bibitem{kruse05}
K.~Kruse, J.~F. Joanny, F.~J{\"u}licher, J.~Prost, and K.~Sekimoto.
\newblock Generic theory of active polar gels: A paradigm for cytoskeletal
  dynamics.
\newblock {\em The European Physical Journal E}, 16(1):5--16, January 2005.

\bibitem{saintillan08}
David Saintillan and Michael~J. Shelley.
\newblock Instabilities and {{Pattern Formation}} in {{Active Particle
  Suspensions}}: {{Kinetic Theory}} and {{Continuum Simulations}}.
\newblock {\em Physical Review Letters}, 100(17):178103, April 2008.

\bibitem{giomi13}
Luca Giomi, Mark~J. Bowick, Xu~Ma, and M.~Cristina Marchetti.
\newblock Defect {{Annihilation}} and {{Proliferation}} in {{Active Nematics}}.
\newblock {\em Physical Review Letters}, 110(22):228101, May 2013.

\bibitem{gao15}
Tong Gao, Robert Blackwell, Matthew~A. Glaser, M.~D. Betterton, and Michael~J.
  Shelley.
\newblock Multiscale modeling and simulation of microtubule--motor-protein
  assemblies.
\newblock {\em Physical Review E}, 92(6):062709, December 2015.

\bibitem{white15b}
D.~White, G.~de~Vries, J.~Martin, and A.~Dawes.
\newblock Microtubule patterning in the presence of moving motor proteins.
\newblock {\em Journal of Theoretical Biology}, 382:81--90, October 2015.

\bibitem{maryshev18}
Ivan Maryshev, Davide Marenduzzo, Andrew~B. Goryachev, and Alexander Morozov.
\newblock Kinetic theory of pattern formation in mixtures of microtubules and
  molecular motors.
\newblock {\em Physical Review E}, 97(2):022412, February 2018.

\bibitem{furthauer19}
Sebastian F{\"u}rthauer, Bezia Lemma, Peter~J. Foster, Stephanie~C.
  {Ems-McClung}, Che-Hang Yu, Claire~E. Walczak, Zvonimir Dogic, Daniel~J.
  Needleman, and Michael~J. Shelley.
\newblock Self-straining of actively crosslinked microtubule networks.
\newblock {\em Nature Physics}, 15(12):1295--1300, December 2019.

\bibitem{ziebert07}
Falko Ziebert, Igor~S. Aranson, and Lev~S. Tsimring.
\newblock Effects of cross-links on motor-mediated filament organization.
\newblock {\em New Journal of Physics}, 9(11):421--421, November 2007.

\bibitem{Giomi2013}
Luca Giomi, Mark~J Bowick, Xu~Ma, and M~Cristina Marchetti.
\newblock {Defect Annihilation and Proliferation in Active Nematics}.
\newblock {\em Physical Review Letters}, 110(22):228101, may 2013.

\bibitem{Lenz2014}
Martin Lenz.
\newblock {Geometrical Origins of Contractility in Disordered Actomyosin
  Networks}.
\newblock {\em Physical Review X}, 4(4):041002, oct 2014.

\bibitem{Kruse2000}
K~Kruse and F~J{\"{u}}licher.
\newblock {Actively Contracting Bundles of Polar Filaments}.
\newblock Technical report, 2000.

\bibitem{Kruse2005}
K.~Kruse, J.~F. Joanny, F.~J{\"{u}}licher, J.~Prost, and K.~Sekimoto.
\newblock {Generic theory of active polar gels: A paradigm for cytoskeletal
  dynamics}.
\newblock {\em European Physical Journal E}, 16(1):5--16, jan 2005.

\bibitem{ahmadi05}
A.~Ahmadi, T.~B. Liverpool, and M.~C. Marchetti.
\newblock Nematic and polar order in active filament solutions.
\newblock {\em Physical Review E}, 72(6):060901, December 2005.

\bibitem{ahmadi06}
Aphrodite Ahmadi, M.~C. Marchetti, and T.~B. Liverpool.
\newblock Hydrodynamics of isotropic and liquid crystalline active polymer
  solutions.
\newblock {\em Physical Review E}, 74(6), December 2006.

\bibitem{Liverpool2005}
T.~B. Liverpool and M.~C. Marchetti.
\newblock {Bridging the microscopic and the hydrodynamic in active filament
  solutions}.
\newblock {\em Europhysics Letters}, 69(5):846--852, mar 2005.

\bibitem{swaminathan11}
S.~Swaminathan, F.~Ziebert, I.~S. Aranson, and D.~Karpeev.
\newblock Motor-{{Mediated Microtubule Self}}-{{Organization}} in {{Dilute}}
  and {{Semi}}-{{Dilute Filament Solutions}}.
\newblock {\em Mathematical Modelling of Natural Phenomena}, 6(1):119--137,
  2011.

\bibitem{gao15a}
Tong Gao, Robert Blackwell, Matthew~A. Glaser, M.~D. Betterton, and Michael~J.
  Shelley.
\newblock Multiscale {{Polar Theory}} of {{Microtubule}} and
  {{Motor}}-{{Protein Assemblies}}.
\newblock {\em Physical Review Letters}, 114(4):048101, January 2015.

\bibitem{gao17}
Tong Gao, Meredith~D. Betterton, An-Sheng Jhang, and Michael~J. Shelley.
\newblock Analytical structure, dynamics, and coarse graining of a kinetic
  model of an active fluid.
\newblock {\em Physical Review Fluids}, 2(9):093302, September 2017.

\bibitem{blackwell16}
Robert Blackwell, Oliver {Sweezy-Schindler}, Christopher Baldwin, Loren~E.
  Hough, Matthew~A. Glaser, and M.~D. Betterton.
\newblock Microscopic origins of anisotropic active stress in motor-driven
  nematic liquid crystals.
\newblock {\em Soft Matter}, 12(10):2676--2687, 2016.

\bibitem{blackwell17a}
Robert Blackwell, Christopher Edelmaier, Oliver {Sweezy-Schindler}, Adam
  Lamson, Zachary~R. Gergely, Eileen O'Toole, Ammon Crapo, Loren~E. Hough,
  J.~Richard McIntosh, Matthew~A. Glaser, and Meredith~D. Betterton.
\newblock Physical determinants of bipolar mitotic spindle assembly and
  stability in fission yeast.
\newblock {\em Science Advances}, 3(1):e1601603, January 2017.

\bibitem{rincon17}
Sergio~A. Rincon, Adam Lamson, Robert Blackwell, Viktoriya Syrovatkina, Vincent
  Fraisier, Anne Paoletti, Meredith~D. Betterton, and Phong~T. Tran.
\newblock Kinesin-5-independent mitotic spindle assembly requires the
  antiparallel microtubule crosslinker {{Ase1}} in fission yeast.
\newblock {\em Nature Communications}, 8:15286, May 2017.

\bibitem{lamson19}
Adam~R. Lamson, Christopher~J. Edelmaier, Matthew~A. Glaser, and Meredith~D.
  Betterton.
\newblock Theory of {{Cytoskeletal Reorganization}} during
  {{Cross}}-{{Linker}}-{{Mediated Mitotic Spindle Assembly}}.
\newblock {\em Biophysical Journal}, 116(9):1719--1731, May 2019.

\bibitem{edelmaier20}
Christopher Edelmaier, Adam~R Lamson, Zachary~R Gergely, Saad Ansari, Robert
  Blackwell, J~Richard McIntosh, Matthew~A Glaser, and Meredith~D Betterton.
\newblock Mechanisms of chromosome biorientation and bipolar spindle assembly
  analyzed by computational modeling.
\newblock {\em eLife}, 9:e48787, February 2020.

\bibitem{wittmann01}
Torsten Wittmann, Anthony Hyman, and Arshad Desai.
\newblock The spindle: A dynamic assembly of microtubules and motors.
\newblock {\em Nature Cell Biology}, 3(1):E28--E34, January 2001.

\bibitem{wollrab19}
Viktoria Wollrab, Julio~M. Belmonte, Lucia Baldauf, Maria Leptin, Fran{\c c}ois
  N{\'e}del{\'e}c, and Gijsje~H. Koenderink.
\newblock Polarity sorting drives remodeling of actin-myosin networks.
\newblock {\em Journal of Cell Science}, 132(4), February 2019.

\bibitem{brangwynne08}
Clifford~P. Brangwynne, Gijsje~H. Koenderink, Frederick~C. MacKintosh, and
  David~A. Weitz.
\newblock Cytoplasmic diffusion: Molecular motors mix it up.
\newblock {\em The Journal of Cell Biology}, 183(4):583--587, November 2008.

\bibitem{tao05}
Yu-Guo Tao, W.~K. den Otter, J.~T. Padding, J.~K.~G. Dhont, and W.~J. Briels.
\newblock Brownian dynamics simulations of the self- and collective rotational
  diffusion coefficients of rigid long thin rods.
\newblock {\em The Journal of Chemical Physics}, 122(24):244903, June 2005.

\bibitem{visscher99}
Koen Visscher, Mark~J. Schnitzer, and Steven~M. Block.
\newblock Single kinesin molecules studied with a molecular force clamp.
\newblock {\em Nature}, 400(6740):184--189, July 1999.

\bibitem{Klumpp2005}
Stefan Klumpp and Reinhard Lipowsky.
\newblock {Cooperative cargo transport by several molecular motors}.
\newblock {\em Proceedings of the National Academy of Sciences of the United
  States of America}, 102(48):17284--17289, nov 2005.

\bibitem{Muller2008}
Melanie~J.I. M{\"{u}}ller, Stefan Klumpp, and Reinhard Lipowsky.
\newblock {Tug-of-war as a cooperative mechanism for bidirectional cargo
  transport by molecular motors}.
\newblock {\em Proceedings of the National Academy of Sciences of the United
  States of America}, 105(12):4609--4614, mar 2008.

\bibitem{Kunwar2011}
Ambarish Kunwar, Suvranta~K. Tripathy, Jing Xu, Michelle~K. Mattson, Preetha
  Anand, Roby Sigua, Michael Vershinin, Richard~J. McKenney, Clare~C. Yu,
  Alexander Mogilner, and Steven~P. Gross.
\newblock {Mechanical stochastic tug-of-war models cannot explain bidirectional
  lipid-droplet transport}.
\newblock {\em Proceedings of the National Academy of Sciences of the United
  States of America}, 108(47):18960--18965, nov 2011.

\bibitem{Bouzat2016}
Sebasti{\'{a}}n Bouzat.
\newblock {Models for microtubule cargo transport coupling the Langevin
  equation to stochastic stepping motor dynamics: Caring about fluctuations}.
\newblock {\em PHYSICAL REVIEW E}, 93:12401, 2016.

\bibitem{Guo2019}
Si~Kao Guo, Xiao~Xuan Shi, Peng~Ye Wang, and Ping Xie.
\newblock {Force dependence of unbinding rate of kinesin motor during its
  processive movement on microtubule}.
\newblock {\em Biophysical Chemistry}, 253:106216, oct 2019.

\bibitem{Arpag2019}
G{\"{o}}ker Arpağ, Stephen~R. Norris, S.~Iman Mousavi, Virupakshi Soppina,
  Kristen~J. Verhey, William~O. Hancock, and Erkan T{\"{u}}zel.
\newblock {Motor Dynamics Underlying Cargo Transport by Pairs of Kinesin-1 and
  Kinesin-3 Motors}.
\newblock {\em Biophysical Journal}, 116(6):1115--1126, mar 2019.

\bibitem{grill05}
Stephan~W. Grill, Karsten Kruse, and Frank J{\"u}licher.
\newblock Theory of {{Mitotic Spindle Oscillations}}.
\newblock {\em Physical Review Letters}, 94(10), March 2005.

\bibitem{blackwell17}
Robert Blackwell, Oliver {Sweezy-Schindler}, Christopher Edelmaier, Zachary~R.
  Gergely, Patrick~J. Flynn, Salvador Montes, Ammon Crapo, Alireza Doostan,
  J.~Richard McIntosh, Matthew~A. Glaser, and Meredith~D. Betterton.
\newblock Contributions of {{Microtubule Dynamic Instability}} and {{Rotational
  Diffusion}} to {{Kinetochore Capture}}.
\newblock {\em Biophysical Journal}, 112(3):552--563, February 2017.

\bibitem{wang11}
Shenshen Wang and Peter~G. Wolynes.
\newblock On the spontaneous collective motion of active matter.
\newblock {\em Proceedings of the National Academy of Sciences},
  108(37):15184--15189, September 2011.

\bibitem{mathijssen16}
A.~J. T.~M. Mathijssen, A.~Doostmohammadi, J.~M. Yeomans, and T.~N. Shendruk.
\newblock Hydrodynamics of micro-swimmers in films.
\newblock {\em Journal of Fluid Mechanics}, 806:35--70, November 2016.

\bibitem{scipy20}
Pauli Virtanen, Ralf Gommers, Travis~E. Oliphant, Matt Haberland, Tyler Reddy,
  David Cournapeau, Evgeni Burovski, Pearu Peterson, Warren {Weckesser},
  Jonathan {Bright}, Stefan~J. {van der Walt}, Matthew {Brett}, Joshua
  {Wilson}, K.~{Jarrod Millman}, Nikolay {Mayorov}, Andrew R.~J. {Nelson}, Eric
  {Jones}, Robert {Kern}, Eric {Larson}, CJ~{Carey}, lhan {Polat}, Yu~{Feng},
  Eric~W. {Moore}, Jake {Vand erPlas}, Denis {Laxalde}, Josef {Perktold},
  Robert {Cimrman}, Ian {Henriksen}, E.~A. {Quintero}, Charles~R {Harris},
  Anne~M. {Archibald}, Antonio~H. {Ribeiro}, Fabian {Pedregosa}, Paul {van
  Mulbregt}, and SciPy 1.~0 {Contributors}.
\newblock Scipy 1.0: Fundamental algorithms for scientific computing in python.
\newblock {\em Nature Methods}, 2020.

\bibitem{hindmarsh83}
A.~C. HINDMARSH.
\newblock {{ODEPACK}}, a systematized collection of {{ODE}} solvers.
\newblock {\em Scientific Computing}, pages 55--64, 1983.

\bibitem{alberts08}
Alberts.
\newblock Molecular biology of the cell, 5th edition by {{B}}. {{Alberts}},
  {{A}}. {{Johnson}}, {{J}}. {{Lewis}}, {{M}}. {{Raff}}, {{K}}. {{Roberts}},
  and {{P}}. {{Walter}}.
\newblock {\em Biochemistry and Molecular Biology Education}, 36(4):317--318,
  2008.

\bibitem{kawaguchi01}
K.~Kawaguchi and S.~Ishiwata.
\newblock Nucleotide-dependent single- to double-headed binding of kinesin.
\newblock {\em Science (New York, N.Y.)}, 291(5504):667--669, January 2001.

\bibitem{kashina96}
A.~S. Kashina, R.~J. Baskin, D.~G. Cole, K.~P. Wedaman, W.~M. Saxton, and J.~M.
  Scholey.
\newblock A bipolar kinesin.
\newblock {\em Nature}, 379(6562):270--272, January 1996.

\bibitem{gerson-gurwitz11}
Adina {Gerson-Gurwitz}, Christina Thiede, Natalia Movshovich, Vladimir Fridman,
  Maria Podolskaya, Tsafi Danieli, Stefan Lak{\"a}mper, Dieter~R. Klopfenstein,
  Christoph~F. Schmidt, and Larisa Gheber.
\newblock Directionality of individual kinesin-5 {{Cin8}} motors is modulated
  by loop 8, ionic strength and microtubule geometry.
\newblock {\em The EMBO journal}, 30(24):4942--4954, November 2011.

\bibitem{valentine06}
Megan~T. Valentine, Polly~M. Fordyce, Troy~C. Krzysiak, Susan~P. Gilbert, and
  Steven~M. Block.
\newblock Individual dimers of the mitotic kinesin motor {{Eg5}} step
  processively and support substantial loads in vitro.
\newblock {\em Nature Cell Biology}, 8(5):470--476, May 2006.

\bibitem{cochran15}
JC~Cochran.
\newblock Kinesin {{Motor Enzymology}}: {{Chemistry}}, {{Structure}}, and
  {{Physics}} of {{Nanoscale Molecular Machines}}.
\newblock {\em Biophysical Reviews}, 7(3):269--299, February 2015.

\bibitem{shimamoto15}
Yuta Shimamoto, Scott Forth, and Tarun~M. Kapoor.
\newblock Measuring {{Pushing}} and {{Braking Forces Generated}} by
  {{Ensembles}} of {{Kinesin}}-5 {{Crosslinking Two Microtubules}}.
\newblock {\em Developmental Cell}, 34(6):669--681, September 2015.

\bibitem{siyyam97}
H.I. Siyyam and M.I. Syam.
\newblock The modified trapezoidal rule for line integrals.
\newblock {\em Journal of Computational and Applied Mathematics}, 84(1):1--14,
  October 1997.

\bibitem{leveque07a}
Randall~J. LeVeque.
\newblock {\em Finite Difference Methods for Ordinary and Partial Differential
  Equations: Steady-State and Time-Dependent Problems}.
\newblock {Society for Industrial and Applied Mathematics}, {Philadelphia, PA},
  2007.

\end{thebibliography}
